\documentclass{aastex631} 
\usepackage{amsmath}	
\usepackage{mathptmx}
\usepackage{graphicx} 

\usepackage{url}

\usepackage[T1]{fontenc}
\usepackage{amssymb}
\usepackage{color}
\usepackage{soul}
\usepackage{hyperref}
\usepackage{subcaption}
\usepackage{booktabs}
\usepackage{float}
\usepackage{afterpage}
\usepackage{diagbox}
\usepackage{tabularx}
\usepackage{pgffor}
\shorttitle{CSST-SL Lens-finding}
\shortauthors{Xu Li et al.}

\begin{document}

\title[CSST-SL Lens-finding]{CSST Strong Lensing Preparation: a Framework for Detecting Strong Lenses in the Multi-color Imaging Survey by the China Survey Space Telescope (CSST)}

\author{Xu Li}
\affiliation{College of Electronic Information and Optical Engineering, Taiyuan University of Technology, Taiyuan, 030024, China}

\author{Ruiqi Sun \footnote{The first two authors contribute equally to this paper.}}
\affiliation{College of Electronic Information and Optical Engineering, Taiyuan University of Technology, Taiyuan, 030024, China}

\author{Jiameng Lv}
\affiliation{College of Electronic Information and Optical Engineering, Taiyuan University of Technology, Taiyuan, 030024, China}

\author[0000-0001-6623-0931]{Peng Jia}
\affiliation{College of Electronic Information and Optical Engineering, Taiyuan University of Technology, Taiyuan, 030024, China}
\email{robinmartin20@gmail.com}

\author[0000-0001-6800-7389]{Nan Li}
\affiliation{National Astronomical Observatories, Beijing, 100101,China}\email{nan.li@nao.cas.cn}

\author{Chengliang Wei}
\affiliation{Purple Mountain Observatory, Chinese Academy of Sciences, Nanjing 210023, China}

\author{Zou Hu}
\affiliation{National Astronomical Observatories, Beijing, 100101,China}

\author{Xinzhong Er}
\affiliation{South-Western Institute for Astronomy Research, Yunnan University, Kunming, Yunnan 65000, P.R. China}

\author{Yun Chen}
\affiliation{Key Laboratory for Computational Astrophysics, National Astronomical Observatories, Chinese Academy of Sciences, Beijing 100101, China}

\author{Zhang Ban}
\affiliation{Changchun Institute of Optics, Fine Mechanics and Physics, Chinese Academy of Sciences, Changchun, 130033, China}
\author{Yuedong Fang}
\affiliation{University Observatory, Faculty of Physics, Ludwig-Maximilians-Universität, Scheinerstr. 1, 81679 Munich, Germany}
\author{Qi Guo}
\affiliation{National Astronomical Observatories, Beijing, 100101,China}
\author{Dezi Liu}
\affiliation{South-Western Institute for Astronomy Research, Yunnan University, Kunming, 650500, China}
\author{Guoliang Li}
\affiliation{Purple Mountain Observatory, Chinese Academy of Sciences, Nanjing 210023, China}
\author{Lin Lin}
\affiliation{Shanghai Astronomical Observatory, Chinese Academy of Sciences, Shanghai 200030, PR China}
\author{Ming Li}
\affiliation{National Astronomical Observatories, Beijing, 100101,China}
\author{Ran Li}
\affiliation{National Astronomical Observatories, Beijing, 100101,China}
\affiliation{Institute for Frontiers in Astronomy and Astrophysics, Beijing Normal University,  Beijing 102206, China}
\affiliation{School of Astronomy and Space Science, University of Chinese Academy of Science, Beijing 100049, China}
\author{Xiaobo Li}
\affiliation{Changchun Institute of Optics, Fine Mechanics and Physics, Chinese Academy of Sciences, Changchun, 130033, China}
\author{Yu Luo}
\affiliation{Purple Mountain Observatory, Chinese Academy of Sciences, Nanjing 210023, China}
\author{Xianmin Meng}
\affiliation{National Astronomical Observatories, Beijing, 100101,China}
\author{Jundan Nie}
\affiliation{National Astronomical Observatories, Beijing, 100101,China}
\author{Zhaoxiang Qi}
\affiliation{Shanghai Astronomical Observatory, Chinese Academy of Sciences, Shanghai 200030, PR China}
\affiliation{School of Astronomy and Space Science, University of Chinese Academy of Science, Beijing 100049, China}
\author{Yisheng Qiu}
\affiliation{Institute for Astronomy, School of Physics, Zhejiang University, Hangzhou 310027, China}
\author{Li Shao}
\affiliation{National Astronomical Observatories, Beijing, 100101,China}
\author{Hao Tian}
\affiliation{National Astronomical Observatories, Beijing, 100101,China}
\author{Lei Wang}
\affiliation{Purple Mountain Observatory, Chinese Academy of Sciences, Nanjing 210023, China}
\author{Wei Wang}
\affiliation{Changchun Institute of Optics, Fine Mechanics and Physics, Chinese Academy of Sciences, Changchun, 130033, China}
\author{Jingtian Xian}
\affiliation{Changchun Institute of Optics, Fine Mechanics and Physics, Chinese Academy of Sciences, Changchun, 130033, China}
\author{Youhua Xu}
\affiliation{National Astronomical Observatories, Beijing, 100101,China}
\author{Tianmeng Zhang}
\affiliation{National Astronomical Observatories, Beijing, 100101,China}
\author{Xin Zhang}
\affiliation{National Astronomical Observatories, Beijing, 100101,China}
\author{Zhimin Zhou}
\affiliation{National Astronomical Observatories, Beijing, 100101,China}



\begin{abstract}
Strong gravitational lensing is a powerful tool for investigating dark matter and dark energy properties. With the advent of large-scale sky surveys, we can discover strong lensing systems on an unprecedented scale, which requires efficient tools to extract them from billions of astronomical objects. The existing mainstream lens-finding tools are based on machine learning algorithms and applied to cut-out-centered galaxies. However, according to the design and survey strategy of optical surveys by CSST, preparing cutouts with multiple bands requires considerable efforts. To overcome these challenges, we have developed a framework based on a hierarchical visual Transformer with a sliding window technique to search for strong lensing systems within entire images. Moreover, given that multi-color images of strong lensing systems can provide insights into their physical characteristics, our framework is specifically crafted to identify strong lensing systems in images with any number of channels. As evaluated using CSST mock data based on an Semi-Analytic Model named CosmoDC2, our framework achieves precision and recall rates of 0.98 and 0.90, respectively. To evaluate the effectiveness of our method in real observations, we have applied it to a subset of images from the DESI Legacy Imaging Surveys and media images from Euclid Early Release Observations. 61 new strong lensing system candidates are discovered by our method. However, we also identified false positives arising primarily from the simplified galaxy morphology assumptions within the simulation. This underscores the practical limitations of our approach while simultaneously highlighting potential avenues for future improvements.
\end{abstract}




\section{Introduction}
Strong Gravitational Lensing represents a significant category of celestial phenomena, occurring as light emitted by distant cosmic entities traverses regions occupied by supermassive objects such as galaxies, galaxy clusters, or black holes \citep{birrer2022time,meneghetti2013arc,treu2010strong,shajib2022strong,treu2022strong,kneib2011cluster,vegetti2023strong}. Strong lensing systems play a vital role in measuring mass distributions, confining cosmological parameters like the Hubble constant and dark energy density, and affords scientists the opportunity to investigate properties of remote celestial objects \citep{bradavc2002b1422+, treu2002internal, auger2010sloan, sonnenfeld2015sl2s}. It furnishes invaluable insights into understanding the early universe and the evolution of the cosmos.\\

Currently, our knowledge of strong lensing systems is limited to a few hundred instances. However, with ongoing and forthcoming surveys \citep{oguri2010gravitationally, jacobs2017finding}, such as the Legacy Survey of Space and Time (LSST) \citep{ivezic2019lsst, jacobs2017finding}, Euclid \citep{2011arXiv1110.3193L}, and the China Space Station Telescope (CSST) \citep{CSSTHu}, we anticipate a substantial increase in the identification of strong lensing systems. The task of detecting them in future surveys presents a formidable challenge. The intricate shapes of images, including arcs and multiple images, often stem from the complex interplay of strong lensing systems influenced by the internal structures of galaxies and their surrounding environments. Furthermore, additional sources of interference, such as the interal structure of galaxies, noise from background and effects brought by the point spread function (PSF), add to the complexity, rendering the detection of strong lensing systems an even more daunting endeavor.\\

As the volume of observation data continue to expand, the manual approach of visually inspecting strong lensing systems has proven insufficient. Therefore, the adoption of appropriate automated algorithms for detection of strong lensing systems become imperative. Previous automated detection methods have adopted various strategies. These include the search for distinctive features of strong lensing systems, such as arcs and rings, the fitting of geometric parameters to quantify the extent of these features, and searching for blue residuals in images extracted from the galaxy \citep{webster1988automated,bacon2000detection,courbin2000exploring,smith2001hubble,lenzen2004automatic, alard2006automated, estrada2007systematic, seidel2007arcfinder, more2012cfhtls, gavazzi2014ringfinder,  jacobs2017finding}. Another approach has involved generating models of potential lensing galaxies and comparing them with real observation data to pinpoint potential lensing occurrences \citep{brault2015extensive, jacobs2017finding}. Certain methods have involved the removal of galaxies from observation images \citep{chan2015chitah,more2016space,jacobs2017finding}, while others have delved into the analysis of colors and shapes of lensed quasars, employing parameterization techniques for subsequent analysis \citep{brault2015extensive,jacobs2017finding}. The aforementioned approaches have indeed discovered a notable amount of strong lensing system candidates. However, these methods necessitate the manual identification of candidates of strong lensing systems, alongside intricate data analysis. Due to the limitations of modern monitors, which can only display images with a finite range of gray scale levels and channels, many of the physical properties associated with strong lensing systems, contained within variations in gray scale or color, cannot be fully utilized in the detection process. Moreover, the elevated computational intricacy associated with these methods renders them unsuitable for processing a huge volume of data.\\

With the development of machine learning techniques, their application within the field of astronomy has increased. Compared to traditional methods, machine learning algorithms exhibit a heightened ability to efficiently manage large datasets and distinguish valuable features from them. Their swift adaptability to new observation data and robust generalization capabilities render them especially fitting for a large volume of astronomical data. In particular, in 2015, researchers introduced a deep neural network model designed to classify galaxy morphology. This model leveraged the translational and rotational symmetries inherent in galaxy images to achieve classification and redshift estimation \citep{dieleman2015rotation, jacobs2017finding}. Within the realm of galaxy evolution studies, \citet{schawinski2017generative} have proposed deep generative models, notably generative adversarial networks (GANs) \citep{goodfellow2014generative,jacobs2017finding}, and devised an innovative deconvolution technique to recover features from SDSS galaxy images.\\

In the realm of strong lensing systems detection, \citet{petrillo2017finding} have conducted an analysis of Kilo Degree Survey (KiDS) data using Convolutional Neural Networks (CNN). Their efforts yielded successful identification of several candidates of strong lensing systems within the KiDS dataset. \citet{metcalf2019strong} have implemented a variety of methods, including visual inspection, arc and ring finders, support vector machines (SVM), and convolutional neural networks (CNN) in detection of strong lensing systems. These methods provided approaches for strong lensing systems finding and classification in Euclid data. \citet{jia2022detection} have introduced a transformer-based deep neural network (DETR) for strong lensing systems detection, which demonstrated noteworthy proficiency in identifying strong lensing systems on the scale of galaxy clusters. The transformer model is designed to concentrate its attention on key regions like distorted images or arcs during the process of strong lensing systems detection. Through its adaptive allocation of attention across different components, the algorithm effectively captures both local and global information, thereby enhancing the efficiency of the detection process.\\

Building upon the methodology introduced in \citet{jia2022detection}, we have further implemented additional efforts. These enhancements incorporate the utilization of a visual transformer employing a sliding window strategy \citep{liu2021swin}. This mechanism divides the image into an array of overlapping sub-windows, treating each sub-window as an input for processing, which addresses the impact of input length restrictions and the imbalance of positive and negative samples on the results. By operating at the window level, the model gains the ability to effectively capture features of diminutive and densely clustered objects, thereby augmenting its capability to detect targets. With these adjustments, our approach is now capable of handling input data of varying lengths, and it results in more equitable feature extraction for a more comprehensive representation of strong lensing phenomena.\\

Furthermore, given the current limited number of detected strong lensing instances and various forms of noise that affect observation images, we have developed an end-to-end pipeline based on the previously discussed detection strategy. This pipeline encompasses three distinct components: image simulation for training data generation, image pre-processing, and target detection. The image simulation component serves to generate a substantial volume of simulated images. These simulated images incorporate prior knowledge about strong lensing systems, which is then utilized to train the detection algorithm. The subsequent image pre-processing step is employed to process real observation images. This involves deconvolution of observation images with predefined PSFs from the telescope and adjusting the grayscales of these images. The image pre-processing component enhances the visibility of celestial objects with low signal-to-noise ratios. For the detection component, the simulated images are used as the training data set. After training, real observation images known to contain strong lensing systems are processed using the image pre-processing algorithm. The outcomes of the image pre-processing algorithm are employed to fine-tune the previously trained detection algorithm. Finally, all real observation images undergo pre-processing before being subjected to the detection algorithm. This sequential process facilitates the identification of strong lensing systems in the observation data.\\

This paper is structured as follows. In Section 2, we elaborate on the methodology employed to generate simulated images featuring strong lensing systems. In this paper, we generate simulated images with both galaxy scale and galaxy cluster scale strong lensing systems, therefore, our method could detect strong lensing systems with both galaxy scale and galaxy cluster scale. In this article, the primary focus is on detecting strong lensing systems at the galaxy scale. In Section 3, we outline the procedures involved in data pre-processing. In Section 4, we introduce the detection algorithm for strong lensing systems, leveraging the attention mechanism. We subsequently evaluate the efficiency of our model using simulated data acquired through the CSST. In Section 5, we establish a comprehensive pipeline for detecting strong lensing systems and assess its performance using public media images from Euclid Early Release Observations and images from the Legacy Imaging Surveys. Finally, in Section 6, we present our conclusions and anticipate potential avenues for future research.\\

\section{The Method to Generate Simulated Data}
\label{sect:Obs}
The China Space Station Telescope (CSST) is a 2-meter space telescope slated to launch in approximately 2025, operating within the same orbit as the China Manned Space Station \citep{CSSTHu}. As a significant scientific project within the Space Application System of the China Manned Space Program, CSST is designed to be a large field of view, covering around 1.1 square degrees, with high spatial resolution of roughly 0.1500 arcseconds at 633 nm for its Survey Camera (SC). This versatile telescope spans multiple wavelength bands, from the near-ultraviolet (NUV) to the near-infrared (NIR), and it will simultaneously conduct photometric and slitless spectral surveys across an extended sky area of 17,500 square degrees.\\

Generating simulated data, as discussed in our previous papers \citep{jia2022detection, jia2023target}, serves a crucial purpose in the detection of celestial objects, particularly when sufficient training data is lacking. Simulated data provide valuable prior information about the targets we aim to detect. These simulations are important in training neural networks, enabling them to swiftly adapt to real observation data through transfer learning. Additionally, we leverage simulated CSST data to make predictions regarding the scientific outcomes of the CSST in the realm of strong lensing detection. The simulation code consists of two key components: the strong lensing simulator and the imaging simulator tailored for the CSST, both of which will be briefly introduced below.\\

We use the PICS strong lensing system simulator without introducing noise or other effects \citep{li2016pics}. Our simulation process, akin to \citet{Madireddy2019}, involves six steps: (1) generating lens and source populations based on statistical properties, (2) constructing mass and light models for foreground lensings, (3) calculating lens deflection fields, (4) creating light profiles for background source galaxies, (5) conducting ray-tracing simulations to generate strong lensing images using deflection fields and source light profiles.  \\

The populations of lenses and sources are built upon an advanced extragalactic catalog \citep{2019Korytov} named \textit{\bf CosmoDC2}, which is a synthetic galaxy catalog to support precision science with the Legacy Survey of Space and Time\footnote{
url{https://rubinobservatory.org/explore/lsst}} (LSST). It covers a $440 deg^2$ of sky area to a redshift of $z=3$ and is complete to a magnitude depth of 28 in the r-band. Various properties, including stellar mass, morphology, spectral energy distributions, broadband filter magnitudes, host halo information, and weak lensing shear, characterize each galaxy in the catalog. The CosmoDC2 has undergone a wide range of observation-based validation tests to ensure consistency with real observations. The official release of the cosmoDC2 dataset and documentation can be found here\footnote{\url{https://portal.nersc.gov/project/lsst/cosmoDC2/_README.html}}. \\

Lenses are modeled as a combination of dark matter halos and galaxies. The dark matter halo follows the Navarro, Frenk \& White (NFW) model, given by \citet{NFW1996}, as shown below:
\begin{equation}
\rho(r)=\frac{\rho_{\mathrm c}}{(r/r_s)
(1+r/r_s)^2}\,,
\label{rho_nfw}
\end{equation}
where $\rho_{\mathrm c}$ is a characteristic density and $r_s$ a scale radius. Its lensing potential is given by \citet{golse2002pseudo},
\begin{equation}
\varphi(x)=2\kappa_s\theta_s^2\,h(x)\,
\label{phi_h}
\end{equation}
with $\kappa_s= \rho_c r_s\Sigma_\mathrm{crit}^{-1}$, $\theta_s = r_s/D_a(z_l)$, and $x=\theta/\theta_s$. Where $D_a(z_l)$ stands for the angular diameter distance from the observer to the lens plane, and 
\begin{equation}
h(x)=
\begin{cases}
\displaystyle{\ln^2{\frac{x}{2}}-\mathrm{arcch}^2
\frac{1}{x}} & (x<1)\\
\displaystyle{\ln^2{\frac{x}{2}}+\arccos^2{\frac{1}{x}}}
 & (x\ge1)
\end{cases}
\end{equation}
To achieve an elliptical lensing potential $\phi_{\epsilon}(x) \equiv \phi(x_{\epsilon})$, \cite{golse2002pseudo} defined $\varepsilon = (1-q^2)/(1+q^2)$ to create the below elliptical coordinate system then substituted $x$ by $x_\varepsilon$.
\begin{equation}
\label{defin_ell}
\left\lbrace
\begin{array}{lcl}
x_{1\epsilon} & = & \sqrt{1-\epsilon} \, x_1 \\
x_{2\epsilon} & = & \sqrt{1+\epsilon} \, x_2 \\
x_\epsilon & = & \sqrt{x_{1\epsilon}^2 +
x_{2\epsilon}^2}\ =\  \sqrt{a_{1\epsilon}x_1^2 +a_{2\epsilon}x_2^2}\\
\phi_\epsilon & = & \arctan \left(x_{2\epsilon} / x_{1\epsilon}\right)
\end{array}
\right.
\end{equation}
Then the deflection angles can be calculated numerically according to the relation $\alpha_1 = \partial \phi_{\epsilon}(x)/\partial x_1$ and $\alpha_2 = \partial \phi_{\epsilon}(x)/\partial x_2$. Therefore, the deflection angles for the dark matter halos are described by $\{x_{nfw,1}, x_{nfw,2}, M_{vir}, c_{vir}, q_{nfw}, \phi_{nfw}, z_l, z_s\}$, where $(x_{nfw,1}, x_{nfw,2})$ is the angular position of the dark matter halo in the field of view, $M_{vir}$ is the virial mass, $c_{vir}$ is the concentration, $q_{nfw}$ and $\phi_{nfw}$ are the axis ratio and positional angle. $z_l$ and $z_s$ represent the redshifts of the lens and source planes, respectively.\\

The density profile of galaxies are modeled as a singular isothermal ellipsoid (SIE), has deflection angles given by \cite{Kormann1994,Keeton2001},
\begin{equation}
    \label{eq:sie-lq-x1}
    \alpha_1 = \frac{\sqrt{q}\,\theta_{\mathrm E}}{\sqrt{(1-q^2)}}\tan^{-1} \left[ \frac{\sqrt{1-q^2}x_1}{\sqrt{q^2 x_1^2 + x_2^2}}\right]\,\,,
\end{equation}
\begin{equation}
    \label{eq:sie-lq-x2}
    \alpha_2 = \frac{\sqrt{q}\,\theta_{\mathrm E}}{\sqrt{(1-q^2)}}\tanh^{-1} \left[ \frac{\sqrt{1-q^2} x_2}{\sqrt{q^2 x_1^2 + x_2^2}}\right]\,\,,
\end{equation}
where $q$ is the minor to major axis ratio and $\theta_{\mathrm E}$ is an effective factor to represent Einstein radius,
\begin{equation}
    \label{eq:sie-re-0}
    \theta_{\mathrm E} = 4 \pi \left( \frac{\sigma_v}{c}\right)^2 \frac{D_{ls}}{D_s}\,\,.
\end{equation}
Therefore, the complete parameter set required by equations~\ref{eq:sie-lq-x1} to equation~\ref{eq:sie-re-0}) is $\{x_1, x_2, \sigma_{v}, q_{gal}, \phi_{gal}, z_l\}$, where $(x_1, x_2)$ is the angular position of the galaxy centers in the field of View, $\sigma_{v}$ is the velocity dispersion of the galaxy, $q_{gal}$ is the axis ratio, $\phi_{gal}$ is the position angle, and $z_l$ is the redshift of the lens plane. The parameters ${x_1, x_2, q_{gal}, \phi_{gal}, z_l}$ are taken directly from the cosmoDC2 catalog.
$\sigma_v$ is derived from the $L-\sigma$ scaling relation from the bright sample of \cite{parker2007masses} given by
\begin{equation}
\sigma_v = 142 \left( \frac{L}{L_{\star}} \right)^{(1/3)}~{\rm km~s}^{-1}\,\, ,
\end{equation}
where, $\log_{10}(L/L_{\star}) = -0.4(mag_r - mag_{r \star})$, and $mag_r$ is the apparent $r$-band magnitude of the galaxy given by the cosmoDC2 catalog.  We use apparent r-band magnitudes ($magr$) from the CosmoDC2 catalog and adopt an evolving $mag_{r \star} = +1.5(z-0.1)-20.44$ along with redshift as per \cite{more2016space} and \cite{Faber2007}. \\

For a given lens, summing all gravitational lensing contribution of its components together, we obtain the total deflection angle $\bf{\alpha_{tot}}$ then the lensing equation $\bf{y} = \bf{x} - \bf{\alpha_{tot}}$. Hence, the strong lensed arcs can be generated by tracing light rays from the lens plane to the source plane, where the source images are drawn according to their positions and light profiles. To ensure prominent lensing features, we require $z_s > z_l+0.5$ for source galaxies and locate one source manually by randomly choosing source positions in regions with lensing magnifications exceeding 20 on the source plane. Galaxies in the light cone, whether lensed or not, are modeled with composite Sersic profiles based on CosmoDC2 parameters. Lensed arcs are rendered with Sersic profiles on ray-traced grids.\\

With the above strategy of generating strong lensing system images, we utilize the CSST simulator, an End-to-End pipeline integrating cosmology simulation, gravitational ray-tracing, optical instruments, and imaging, to generate mock observation images\footnote{\url{https://csst-tb.bao.ac.cn/code/csst_sim/csst-simulation}}. In this simulator, galaxies consist of bulges and disks, which are modeled with composite Sersic profiles based on CosmoDC2 parameters, and they are convolved with simulated PSFs for CSST. These galaxies are then imaged onto detectors with dimensions of $9216\times 9232$ pixels with pixelscale of 0.074 arcseconds. Then we choose strongly lensed arcs in three steps: (1) calculate the reference Einstein Radii $\theta_{E, ref}$ of all galaxies and galaxy clusters in the above field of view by setting $z_s = 10.0$; (2) remove the deflectors with $\theta_{E, ref}) < 0.2 "$; (3) randomly choose $1\%$ of the rest deflectors as lenses. We then generate strongly lensed arcs for the selected lenses as described in the above paragraphs, and the pixelized images of the strongly lensed arcs are passed into the CSST-image-simulator afterwards. Subsequently, both galaxies and strong strong lensing systems are imaged together onto the detectors.\\

Rather than relying on parametric models, the PSF is derived from an optical design model. To produce a comprehensive set of realistic PSFs that consider the impact brought by the optical system, an optical emulator is created to simulate high-fidelity PSFs for CSST. This optical emulator for CSST comprises six distinct modules, each simulating optical aberrations stemming from mirror surface roughness, fabrication errors, CCD assembly errors, gravitational distortions, and thermal distortions. Additionally, the simulated PSF incorporates two dynamic errors arising from micro-vibrations and image stabilization. We have further introduced various sources of noise into the simulated CSST images, encompassing shot noise, sky background, and detector effects. To accomplish this, we have employed Galsim \citep{rowe2015galsim} to simulate photon generation from a given galaxy, accounting for the throughputs of the CSST system, including mirror efficiency, filter transmission, and detector quantum efficiency. We have also introduced Poisson noise originating from both the sky background and the dark current of the CCD detector. Specifically, we have set the i-band background level to 0.212 $e^-$/pixel/s, with a dark current of 0.02 $e^-$/pixel/s, resulting in an average of approximately 35 $\rm e^-$/pixel in a 150s exposure. Furthermore, we have incorporated read noise using a Gaussian distribution with a standard deviation of approximately 5.0 $e^-$/pixel. In simulating the creation of mock galaxy images on the detector, we have also accounted for bias and applied gain factor of the detector.\\

\section{The Data Pre-processing Method}
\label{sect:data}
Within this section, we establish the data pre-processing method for images captured by the CSST. It is worth noting that these steps can be adjusted, or even omitted, when they are used to process observation data obtained by other telescopes. Moreover, should the need arise, parameters within these processes can be readily fine-tuned to adapt with images obtained by other surveys.\\

\subsection{The Image Cropping Step}
\label{subsec:ImgCro}
Given that images acquired by the CSST are relatively large, with size of $9232\times 9216$ pixels, we find it is necessary to crop them into smaller sections, each with size of $1000\times 1000$ pixels. This step is essential to align with hardware constraints, as the detection algorithm needs approximately 14.88 GB GPU memory for an image of dimensions $1000\times 1000$ pixels. Given that this study uses an RTX 3090 Ti GPU with a maximum memory capacity of 24 GB, this image size represents the upper limit of what the GPU can accommodate. Moreover, to ensure that images of strong lensing systems remain intact and undivided, we have introduced overlapping segments, with each segment featuring an overlap region of 80 pixels. This approach results in the generation of 100 smaller images following the cropping of a single original larger image. It is worth noting that the image cropping step can be circumvented if access to a GPU with greater memory capacity is available.\\

\subsection{The Image Deconvolution Step}
\label{subsec:Deconv}
\begin{figure*}
\centering
\includegraphics[width=0.9\textwidth]{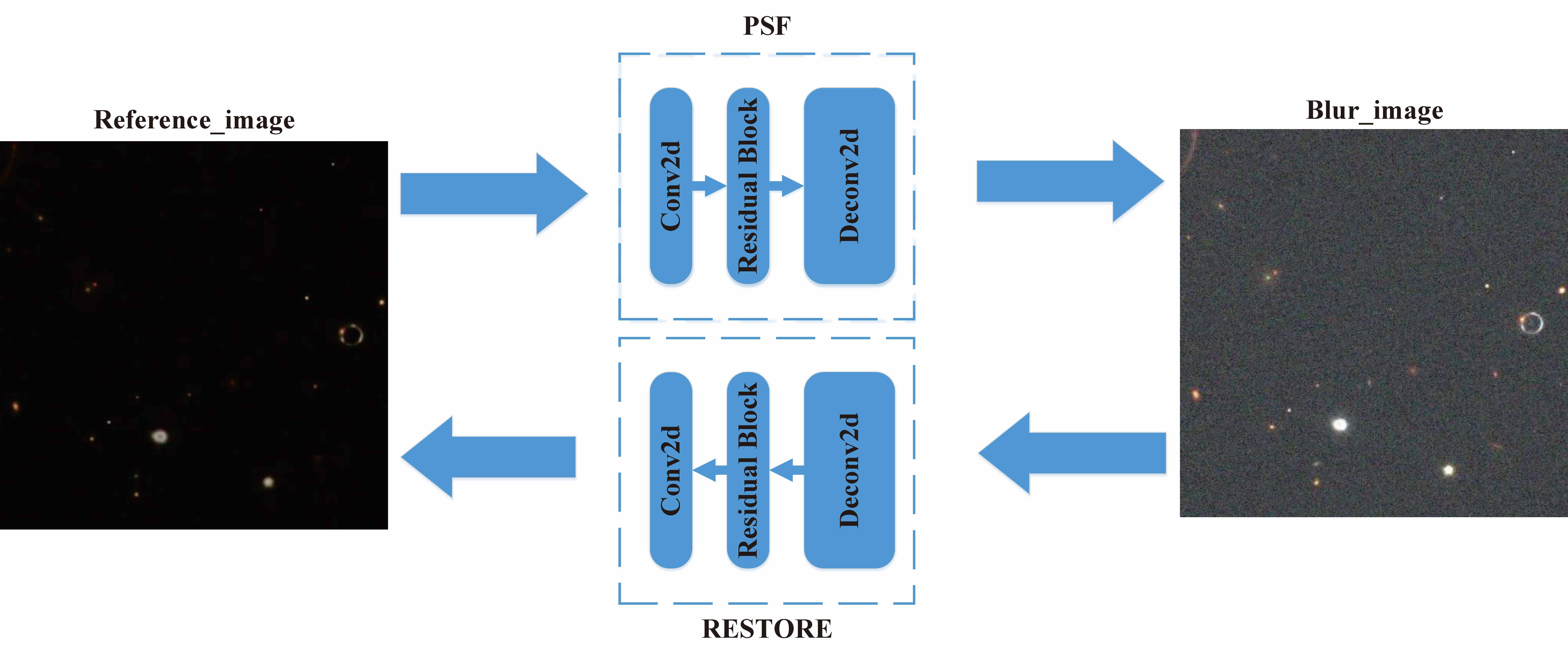}
\caption{The flowchart of image restoration step, which contains an image restoration neural network (RESTORE) and an image blurring neural network (PSF).}
\label{Fig1}
\end{figure*}

Astronomical imaging faces inherent constraints in angular resolution, primarily attributed to factors like atmospheric turbulence, diffraction, instrumental effects, and others. This limitation is rooted in the mathematical equivalence of Fraunhofer diffraction to a Fourier transform, indicating the absence of signal on Fourier scales smaller than the diffraction limit. Consequently, when the PSF sets the diffraction-limited optics of the telescope, resulting in zero signal-to-noise ratio on scales smaller than the PSF, the image information becomes incomplete. Additional factors, such as jitter, may exacerbate the performance degradation. Leveraging advancements in machine learning, deconvolution neural networks have been employed to augment and suppress noise in astronomical images across various wavelengths, with the goal of restoring diffraction-limited performance \citet{Lauritsen2021}. In this study, we assume that we can obtain PSFs of the CSST and these images can be used to generate simulated blurred images. These blurred images and the original images can be used as the training set for deep learning based image restoration algorithms. We employ the PSF-NET method for image deconvolution, as presented by \citet{jia2020psf}. The PSF-NET contains two neural networks, a PSF neural network and a restoration neural network. The PSF neural network models the PSF, facilitating the conversion of high-resolution images into blurred counterparts. The restoration neural network, named RESTORE, encapsulates the deconvolution algorithm, thereby converting blurred images back to high-resolution forms. Both of these neural networks include residual blocks, in addition to multiple convolutional and transposed convolutional blocks, as illustrated in Figure~\ref{Fig1}.\\

During the training phase, both the PSF neural network and the RESTORE neural network are trained. This approach enhances training efficiency while mitigating the risk of overfitting. Subsequent to training, the RESTORE neural network could be used directly to restore blurred images. In accordance with the functionalities of these two networks, we adjust the loss function of these neural networks as depicted in Equation \ref{equation1}, following the method proposed in \citep{lv2022general, jia2024image}. Here, $L_{idet}$ signifies the identity loss function, $L_{rec}$ denotes the cyclic loss function, and $L_{fl}$ stands for the focal frequency loss function.
\begin{equation}
\label{equation1}
\begin{aligned}
	Loss =L_{idet}+L_{rec}+L_{fl},\\
	L_{idet} = \Vert PSF(Img_{org})-Img_{blur} \Vert_2\\ 
	+  \Vert Restore(Img_{blur})-Img_{org} \Vert_2,\\
	L_{rec} = \Vert Restore(PSF(Img_{org}))-Img_{org} \Vert_2\\ 
	+  \Vert PSF(Restore(Img_{blur}))-Img_{blur} \Vert_2,\\
	L_{ffl} = W \times L_{fl}.
\end{aligned}
\end{equation}
In the above equation, both $L_{idet}$ and $L_{rec}$ are used to minimize the mean square error that exists between the original image and the image restored by the neural network. The PSF neural network and the RESTORE neural network share an encoder-decoder structure. However, the decoder section encompasses several convolutional and upsampling layers. These layers introduce gaps in the restored image, leading to loss of information and the introduction of artifacts. Such issues can disrupt the accurate interpretation and analysis of the image. To enhance the efficacy of our neural network, we introduce a focal frequency loss, denoted as $L_{ffl}$ \citep{jiang2021focal}. This particular loss function contributes to the improvement of neural network performance by incorporating regularization weights into the density of the power spectrum, W. The loss of focal frequency, $L_{ffl}$, is equivalent to the mean square error computed in the spatial frequency domain between the restored image and the original image, as shown in Equation \ref{equation2}. In this equation, $FFT$ denotes the Fast Fourier Transformation, and $\alpha$ corresponds to the regularization parameter, which is assigned a value of 1.

\begin{equation}
\label{equation2}
\begin{aligned}
	W = \left |FFT(Img_{org}) - FFT(Restore(Img_{blur})) \right |^\alpha,\\
	L_{fl} =  \Vert FFT(Img_{org}) - FFT(Restore(Img_{blur})) \Vert^2.
\end{aligned}
\end{equation}

\begin{figure*}
\centering
\subcaptionbox{\label{subfig:a}}{
\includegraphics[width=0.2\textwidth]{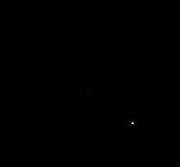}
}
\subcaptionbox{\label{subfig:b}}{
\includegraphics[width=0.2\textwidth]{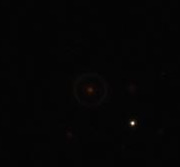}
}
\subcaptionbox{\label{subfig:c}}{
\includegraphics[width=0.2\textwidth]{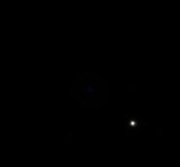}
}
\subcaptionbox{\label{subfig:d}}{
\includegraphics[width=0.2\textwidth]{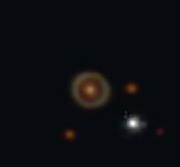}
}
\caption{We chose the r, g, and i bands to generate a color image. Figure (a) shows the raw data, Figure (b) shows the data after the asinh transformation, Figure (c) displays the restoration results and Figure (d) shows the restored image after the asinh grayscale transformation. As depicted in these figures, the incorporation of image restoration and grayscale transformation steps significantly enhances the visibility of image details.}
\label{Fig2}.
\end{figure*}

In the provided equation, $L_{idet}$ and $L_{rec}$ are employed to minimize the mean squared error (MSE) between the original image and the image restored by the neural network. Adhering to the image restoration principles discussed earlier, we have performed restoration on the simulated data that contained noise. Following this, we can compare the restored image with the original noisy after applying the asinh transformation. The restoration results are visualized in Figure~\ref{Fig2}. As shown in this figure, the quality of images have been imporved with our method. We use the peak signal-to-noise ratio (PSNR) defined in \citet{wang2004image, xu2014deep} to evaluate the quality of these images. In this case, the PSNR has been improved from 48.22 to 51.77 by our method. This highlights how our method can effectively improve the signal-to-noise ratio of faint celestial targets.\\

\subsection{The Grayscale Transformation Step}
\label{subsec:grayscale}
In contrast to typical images, astronomical images frequently encompass celestial objects with significantly wider grayscale ranges. Consequently, certain images exhibiting a low signal-to-noise ratio might appear similarly to the background when compared to the overall grayscale. While deep neural networks are capable of acquiring non-linear mapping functions, it might be an inefficient utilization of computational resources if we can establish straightforward transformation functions. A grayscale transformation function is one example of such functions. The asinh function, introduced by \citet{lupton2004preparing}, serves to enhance the signal-to-noise ratio of faint targets. This function is defined by Equation \ref{equation3},
\begin{equation}
asinh(x)=ln(x + sqrt(x^2 + 1))
\label{equation3}
\end{equation}
where $x$ is a real number. When the value of $x$ is relatively large, $asinh(x)$ can be approximated as $ln(x)$. When the value of $x$ is relatively small, $asinh(x)$ can be approximated as $x$.\\

When handling these images, our initial step involves computing the maximum and minimum values within the provided image to establish the data range. This range is then used to derive the scaling factor for the data. It is important to highlight that, when analyzing processed images using our method where the flux has been calibrated across different bands, we can utilize the maximum and minimum values. However, if dealing with real observation data, it is crucial to meticulously choose suitable values based on the photometry zero point and the full well depth of the observed images. The equation to calculate this scaling factor in this paper is illustrated in Equation \ref{equation4},
\begin{equation}
scale\_factor = asinh(0.5 \times \frac{(max\_value - min\_value) }{max\_value}),
\label{equation4}
\end{equation}
where $max\_value$ and $min\_value$ respectively represent the maximum and minimum values of the image. After calculating the scaling factor, the data can be subjected to an asinh transformation, defined in equation \ref{equation5},
\begin{equation}
asinh\_value = asinh(\frac{0.5 \times (data\_value - min\_value)}{2 \times scale\_factor}),
\label{equation5}
\end{equation}
where $data\_value$ corresponds to the original data value, $asinh\_value$ signifies the value after undergoing the asinh transformation, while $min\_value$ and $scale\_factor$ respectively denote the minimum value and the scaling factor determined during the scaling factor calculation. The figure illustrates the image following the grayscale transformation process. As shown in figure~\ref{Fig2} (c), celestial objects with a low signal-to-noise ratio become distinctly visible.\\

\section{The Strong Lensing System Detection Algorithm Based on the Swin Transformer}
\label{sect:switrans}
\subsection{The Structure of the Strong Lensing System Detection Algorithm}
The algorithm proposed in this paper is built upon the foundation of the transformer model. The transformer, originally applied extensively in the realm of Natural Language Processing (NLP), has yielded favorable outcomes in numerous NLP tasks \citep{vaswani2017attention}. Building on these successes, the transformer architecture has been extended to tasks involving image processing. However, in contemporary transformer-based neural networks, tokens are confined to fixed sizes \citep{vaswani2017attention, dosovitskiy2020image, liu2021swin}, thus restricting the input to a fixed number of dimensions. This presents a challenge when it comes to images, which inherently possess thousands of dimensions and requires precise predictions at the pixel level. As a consequence, the transformer faces difficulties in effectively handling high-resolution images, largely due to the quadratic increase in computational complexity with respect to image size.\\

To surmount these challenges, we implement an enhancement to the transformer architecture by incorporating the Swin Transformer model as a core feature extraction module \citep{liu2021swin, jia2023deep}. The Swin Transformer has two noteworthy advantages. Firstly, its hierarchical structure aligns seamlessly with the feature pyramid network \citep{lin2017feature, liu2021swin}) or the U-Net \citep{ronneberger2015u, liu2021swin}, enhancing its compatibility with such frameworks. Secondly, it introduces the concept of moving windows, where self-attention calculations are confined to a local window, as opposed to global attention calculations \citep{liu2021swin}. This architectural choice significantly mitigates model complexity, reducing the quadratic complexity ($N^2$) to linear complexity. These attributes collectively render the Swin Transformer an adaptable feature extraction model, well-suited for a range of visual tasks.\\

\begin{figure*}
\centering
\includegraphics[width=0.96\textwidth, angle=0]{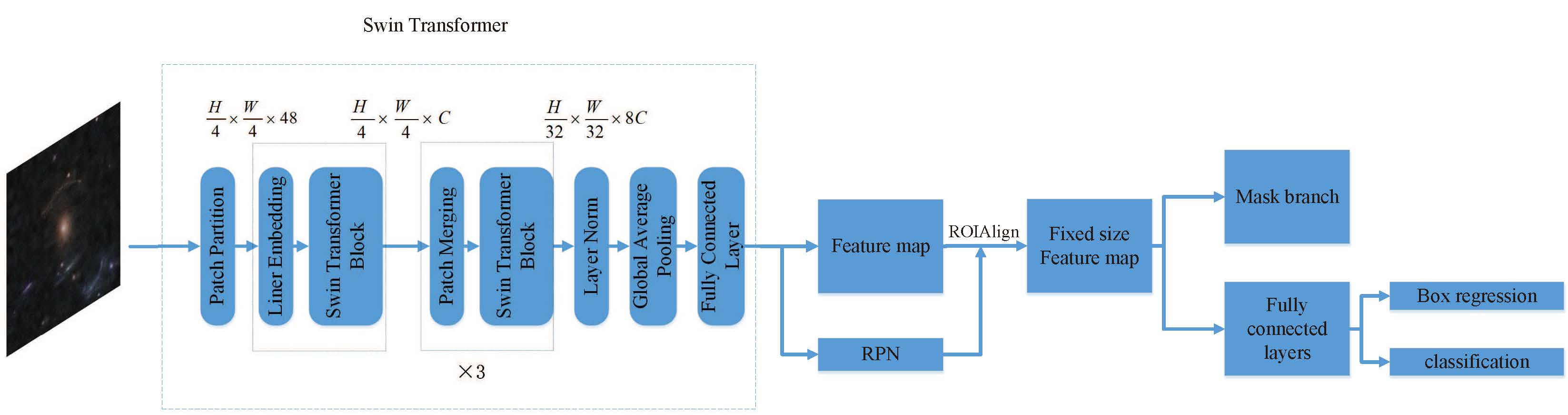}
\caption{As depicted in this figure, our method initiates the feature extraction process using the Swin Transformer. These extracted features are subsequently compiled into a feature map, which is then utilized for box regression and classification.}
\label{Fig3}
\end{figure*}

The architecture of the model is shown in Figure~\ref{Fig3}. Here is an overview of its structure: The initial step involves feeding preprocessed images into the Swin Transformer, which serves as the feature extractor. Subsequently, the outputs from the Swin Transformer are directed into the RPN network for binary classification and regression of positions. Regions potentially containing strong lensing systems are singled out, and using the ROIAlign layer, the feature map is cropped and aligned to generate fixed-size feature tensors. These tensors are then independently transmitted to the classification branch, bounding box regression branch, and mask segmentation branch, each responsible for different aspects of the object detection task. By amalgamating the outputs from these three branches, the network can simultaneously provide information about comprehensive insights regarding the position and the mask of each detected strong lensing systems.\\

\begin{figure*}
\centering
\includegraphics[width=0.9\textwidth, angle=0]{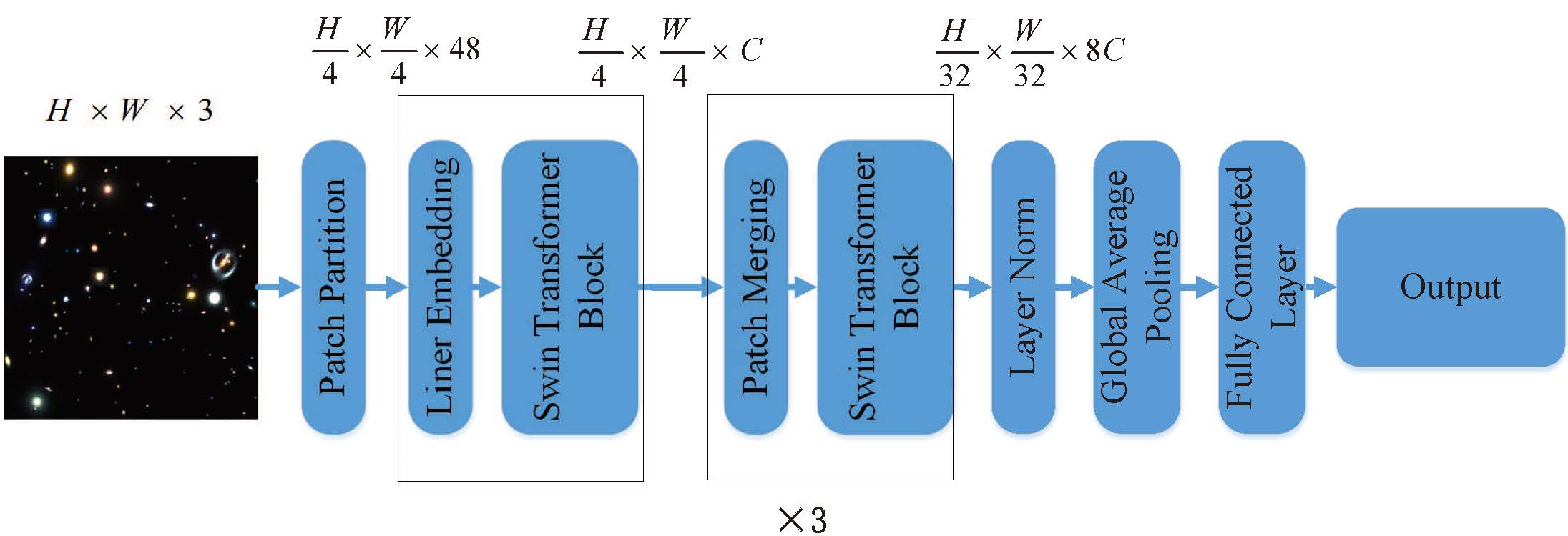}
\caption{The specific structure and data flow of the Swin Transformer. The input image undergoes linear embedding, patch merging, Layer Norm, global pooling, and fully connected layers to obtain the final output.}
\label{Fig4}
\end{figure*}

Figure \ref{Fig4} shows the precise architecture and data flow of the Swin Transformer. An image, sized $H \times W \times C$, is divided into $4 \times 4$ patches, effectively transitioning from pixels to patches as the smallest unit of representation. This procedure generates a feature map of $\frac{H}{4} \times \frac{W}{4} \times (C \times 4 \times 4)$. This feature map undergoes three distinct stages. In the first stage, a linear embedding transforms the feature map from $\frac{H}{4} \times \frac{W}{4} \times (C \times 4 \times 4)$ to $\frac{H}{4} \times \frac{W}{4} \times Channel$. In the subsequent three stages, patch merging is executed, performing operations subsequent to the amalgamation of neighboring patches. For the classification networks, the final output is obtained by connecting a Layer Norm layer, a global pooling layer, and a fully connected layer.\\

\begin{figure*}
\centering
\includegraphics[width=0.9\textwidth, angle=0]{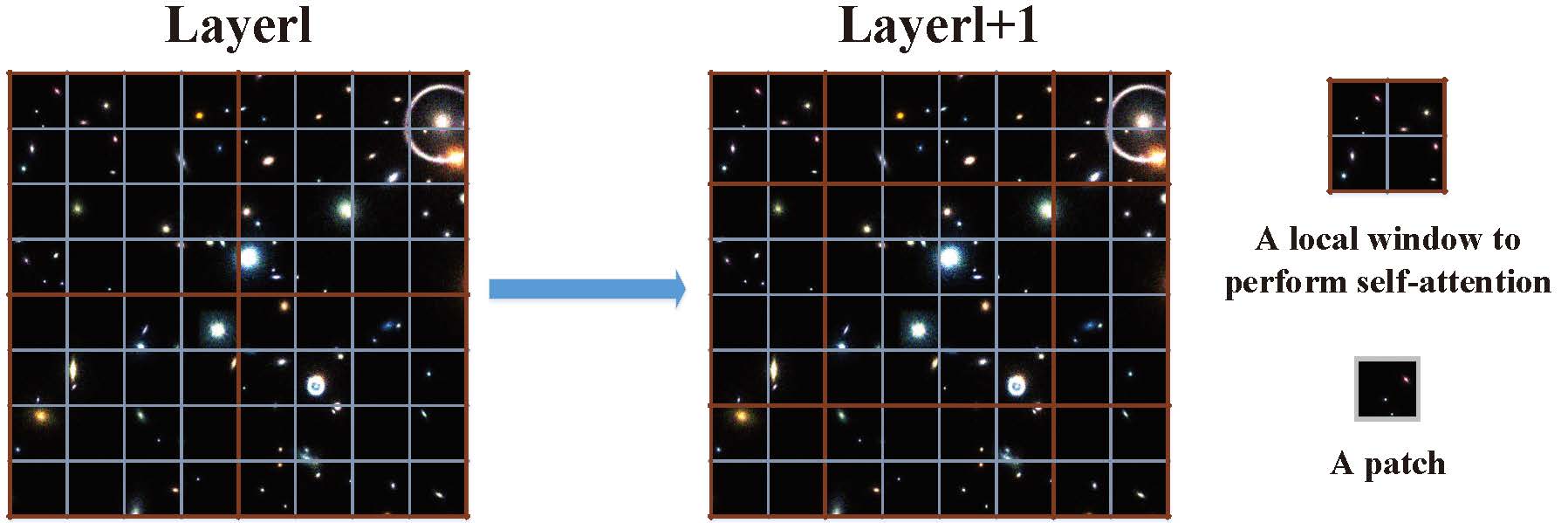}
\caption{The diagram illustrates the computation diagram of self-attention with a shifting window. In Layerl, each window is averaged, and the self-attention is calculated independently for each window. In Layerl+1, the windows are shifted systematically, generating new windows that maintain connectivity between the windows when calculating attention.}
\label{Fig5}
\end{figure*}

In each stage, the initial input undergoes downsampling via a Patch Merging layer. This layer groups neighboring pixels into patches, which are subsequently combined to form four feature maps. These four feature maps are then merged in the depth dimension. Following this, a LayerNorm and fully connected layer are employed to linearly adjust the depth of the feature map, halving its original size to $\frac{C}{2}$. After traversing the Patch Merging layer, the height and width of feature maps are halved, while its depth is doubled. To manage high-resolution images with many tokens more effectively, the Swin Transformer implements a technique known as local window self-attention. This approach reduces computational complexity while maintaining commendable outcomes. In essence, the model concentrates on local regions at a given moment, rather than attempting to process the entire image concurrently. This method facilitates a step-by-step image processing approach, making it feasible to handle a lot of images without overwhelming computational resources. To ensure coherence among windows and boost efficiency, the Swin Transformer adopts a strategy called "shifted window partitioning". This strategy mitigates computational load while ensuring interconnections among the windows, as shown in Figure~\ref{Fig5}.\\

\begin{figure*}
\centering
\includegraphics[width=0.95\textwidth, angle=0]{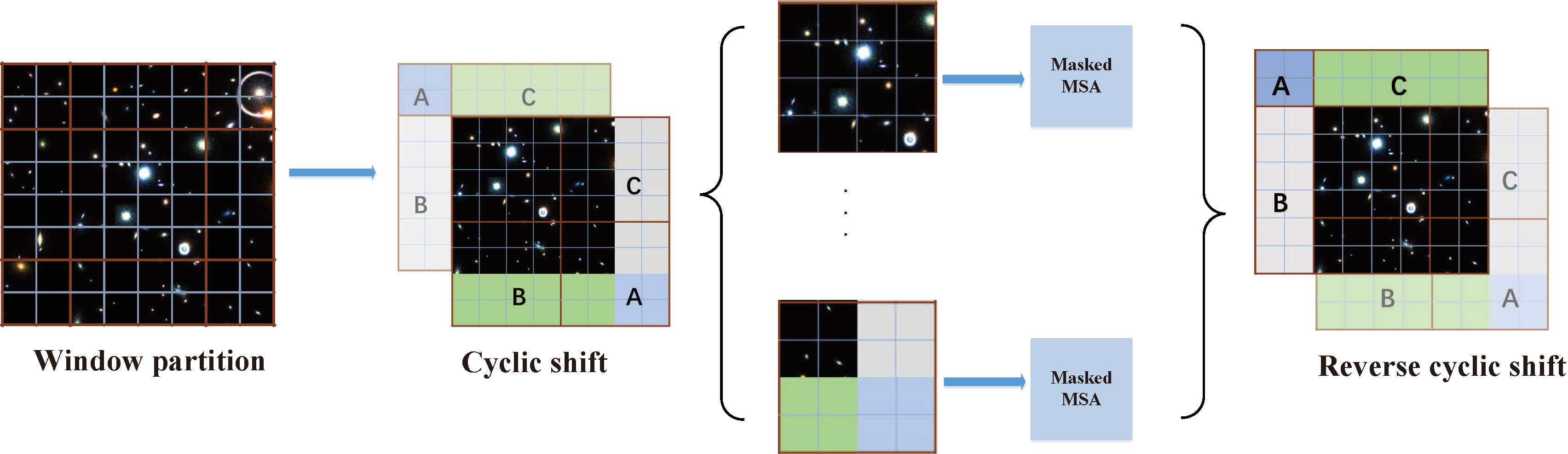}
\caption{The figure illustrates an efficient batch processing approach used for self-attention in the context of shifting window partitioning.}
\label{Fig6}
\end{figure*}

As depicted in Figure \ref{Fig5}, $Layerl$ contains only 4 windows, while $Layerl+1$ encompasses 9 layers. Multi-Scale Analysis (MSA) is executed within each of these windows. To alleviate computational burden, an Efficient Batch Computation approach is employed, utilizing shifted configurations for resolution. This is shown in Figure \ref{Fig6}. The method involves filling relatively small windows to dimensions of $M \times M$ through a shifting operation, while masking the areas that are filled with attention. In Figure \ref{Fig6}, after cyclically shifting to the upper left, the batch window encompasses several sub-windows that are not adjacent in the feature map. The number of batch windows aligns with the window partition count. During the computation step, self-attention is applied solely to each sub-window, with a few others being masked out. After the computation step, the sub-window is repositioned to its original orientation. In simpler terms, this method enlarges windows by shifting them and disregards specific values in the process. This approach yields a batch window composed of disconnected sub-windows. Subsequently, self-attention is carried out individually within each sub-window, and once calculations conclude, the original arrangement is restored.\\

\subsection{Training of the Strong Lensing System Detection Algorithm}
Due to the large number of parameters of the Swin Transformer, utilizing random weights for neural network training would consume substantial computational resources. Therefore, we initialize the network with the weights provided by K-H-Ismai \footnote{\url{https://github.com/SwinTransformer/Swin-Transformer-Object-Detection\#mask-r-cnn}}. As described in Section~\ref{sect:Obs}, the simulated data serves as the training and validation datasets for the neural network. Our simulation assumes a galaxy density of 55.6 galaxies per arcmin$^2$ detectable by the CSST, resulting in an expected 1.4 strong lensing systems per arcmin$^2$. This galaxy density aligns with the Stage IV galaxy surveys. The training dataset consists of 3,000 images, each with dimensions of 1000 x 1000 pixels. Training and validation sets are randomly split with a 7:3 ratio. It is worth noting that if the model is intended for application with data from other sky survey projects, utilizing images generated by specific simulation codes or real observation data for fine-tuning would be necessary, given the specialized nature of the CSST simulation code.\\

\begin{figure}
\centering
\includegraphics[width=0.4\textwidth]{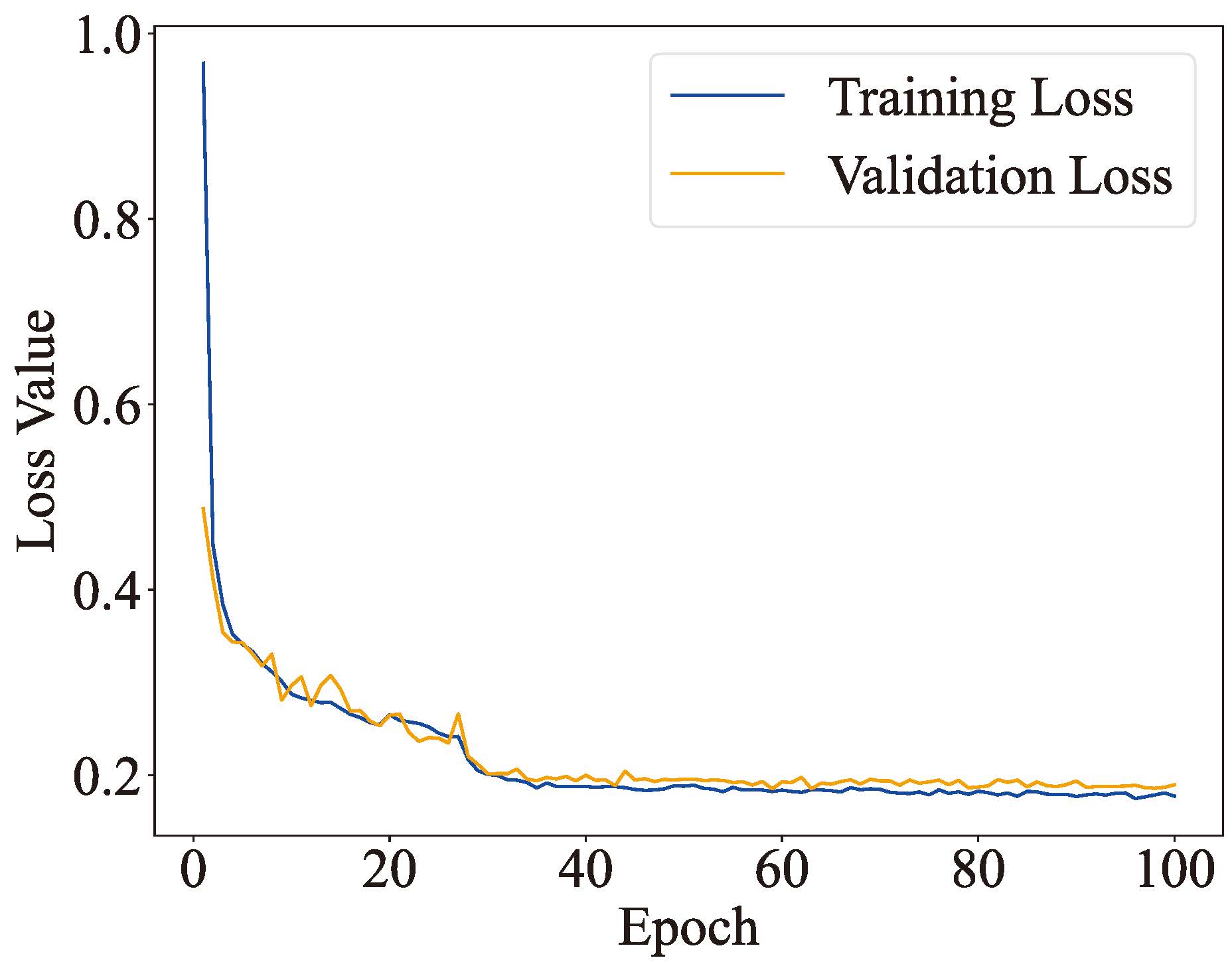}
\caption{This figure illustrates the learning curve observed throughout the entire training and validation process. We can find that the model achieves convergence at approximately 80 epochs.}
\label{Fig7}
\end{figure}

We generate labels for these images using the following approach. Images of strong lensing systems are enclosed with bounding boxes, and we establish these boxes using center coordinates and a fixed size. Since images of strong lensing systems often exhibit varying sizes, we opt for bounding boxes with a fixed dimension of $50 \times 50$ pixels. This strategy enables precise object localization, ensuring that a broader range of objects is captured while minimizing the inclusion of extraneous background details. Consequently, this approach enhances the efficacy of training the model. During the training stage, the learning rate for the Swin Transformer is set as 0.001. As we train the neural network, a consistent decrease in loss could be observed. After 100 epochs of training, which cost around 72 hours in a computer with one RTX 3090 Ti GPU, the loss stops decreasing, which shows that the model is convergence. The learning curve is depicted in Figure \ref{Fig7}.\\

\section{Performance Evaluation of the Pipeline}
\label{sect:discussion}
In this section, we will assess the performance of the pipeline using simulated CSST data and explore the potential scientific insights the CSST could provide for studying strong lensing. Additionally, we will demonstrate the essential nature of various stages within our pipeline.\\

\subsection{Evaluation Metrics of the Detection Results}
In object detection tasks, the accuracy of position prediction is often evaluated using a metric known as Intersection over Union ($IOU$) ratio. This ratio quantifies the degree of overlap between the predicted bounding box and the actual (ground truth) box. Since the position and size of prediction results are both important for object detection tasks, we calculate the $IOU$ value between our prediction and the ground truth box. A higher $IOU$ value indicates more precise predictions. The $IOU$ calculation is defined by Equation \ref{equation7},

\begin{equation}
IOU= \frac{Intersection}{Union},
\label{equation7}
\end{equation}
where $Intersection$ pertains to the overlap region between the predicted bounding box and the actual (ground truth) bounding box, while $Union$ encompasses the entirety of the space covered by both the predicted bounding box and the ground truth bounding box. In simpler terms, $Intersection$ corresponds to the shared area and $Union$ represents the total area covered by both boxes. When evaluating object detection results using the IOU metric, we need to select a specific threshold. If the value $IOU$ between the predicted box and the ground truth box exceeds this threshold, we consider the predicted box to have successfully detected the object, and the model classifies it as true positive ($TP$). On the other hand, if the $IOU$ value falls below the threshold, the predicted box is considered an unsuccessful detection. In this case, the model designates it as either a false negative ($FN$) if it failed to detect a true object or a false positive ($FP$) if it mistakenly identified a non-object region as an object. This threshold-based approach helps us evaluate the performance of the model in distinguishing between true positives, false negatives, and false positives.\\

In order to evaluate the performance of our model, we compute two crucial metrics: precision and recall. Precision informs us about the ratio of accurately classified positive results relative to all positive results, and its calculation is based on the following equation \ref{equation8},
\begin{equation}
Precision= \frac{TP}{TP+FP}.
\label{equation8}
\end{equation}
\indent 
The recall rate, often referred to as the "detection rate," defines the percentage of accurately recognized bounding boxes out of all the actual positive boxes. Its computation is determined by the formula depicted in Equation \ref{equation9},
\begin{equation}
Recall= \frac{TP}{TP+FN}.
\label{equation9}
\end{equation}
\indent 
The $Precision$ and $Recall$ are interconnected metrics with a reciprocal relationship. As we elevate the $IOU$ threshold, both $Precision$ and $Recall$ typically experience a decrease. This signifies that the model becomes more precise in recognizing true positives, yet it could overlook certain positive instances, resulting in a decline in $Recall$. Conversely, by lowering the threshold, the $Recall$ can increase as the model becomes more attuned to positive instances. However, this improvement in $Recall$ comes with a trade-off: the $Precision$ may decrease, potentially leading to more false positive predictions by the model.\\

For an accurate model evaluation, achieving a balance between precision and recall is essential, and this balance can be attained by determining an optimal threshold. In this way, we can achieve a model that correctly identifies positive samples while minimizing the occurrence of false positives. To balance between these parameters, we need to use the Precision-Recall (PR) curve. The PR curve illustrates the interplay between precision and recall. Precision indicates the correctness of positive predictions, while recall gauges the ability of the model to detect all actual positive instances. On the PR curve, we plot recall on the x-axis and precision on the y-axis. By examining this curve, we can make informed decisions about the  performance of the model and select the threshold that aligns with our specific requirements. Typically, the closer the PR curve resides to the upper-right corner, the better the performance is. To quantitatively measure the trade-off between precision and recall for an individual class, we compute the Average Precision (AP), which involves calculating the area under the Precision-Recall curve. Additionally, the mean Average Precision (mAP) represents the average AP value across all classes, offering an assessment of the overall model performance. In this paper, we employ PR curves and mAP to compare the performance of different trained models, aiding in our comprehensive evaluation of their efficacy.\\

\subsection{Performance Evaluation of the Detection Algorithm}
\label{subsect:orgPer}
\begin{figure}
\centering
\subcaptionbox{\label{subfig:a}}{
\includegraphics[width=0.48\textwidth]{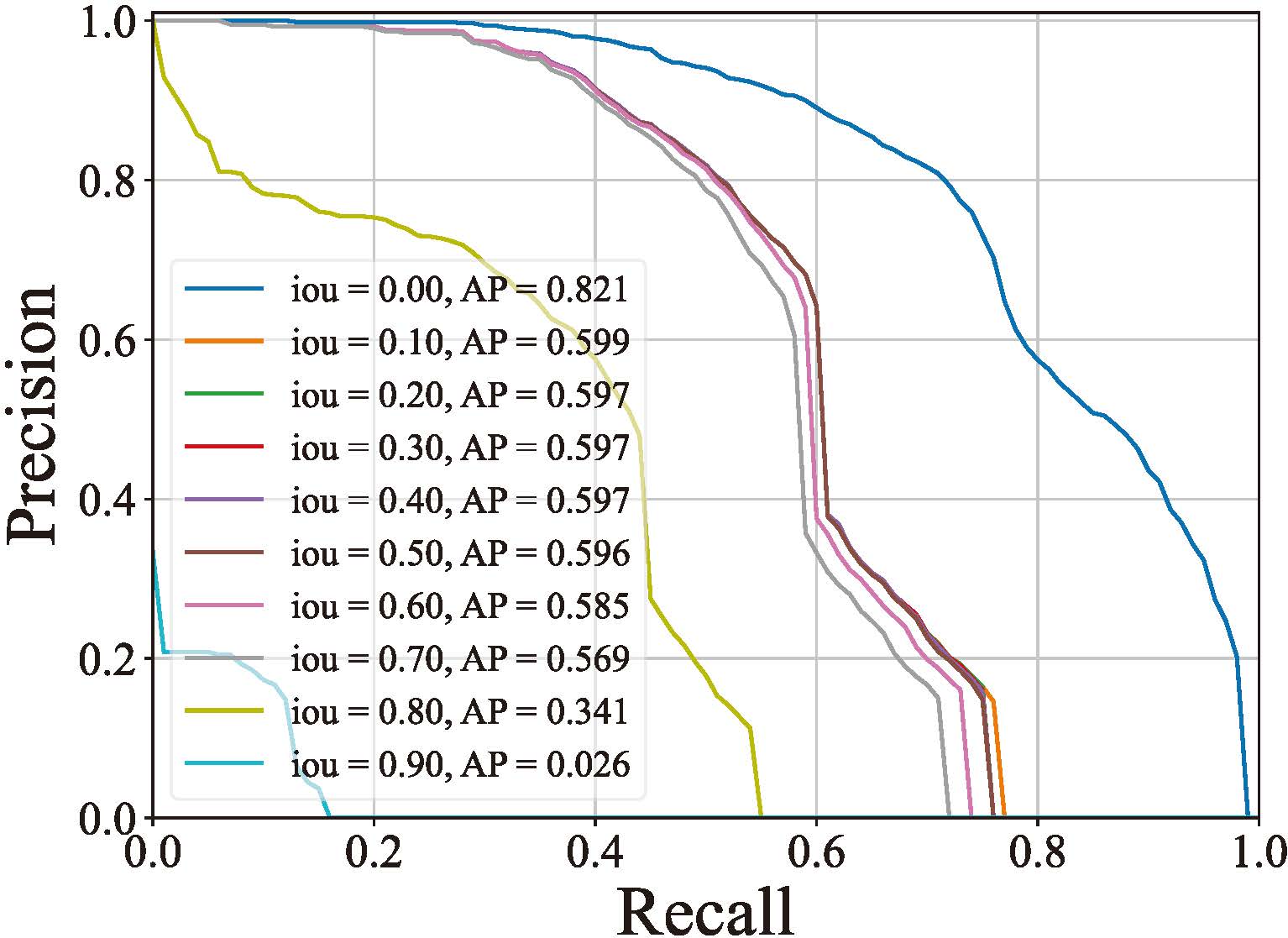}
}
\subcaptionbox{\label{subfig:b}}{
\includegraphics[width=0.48\textwidth]{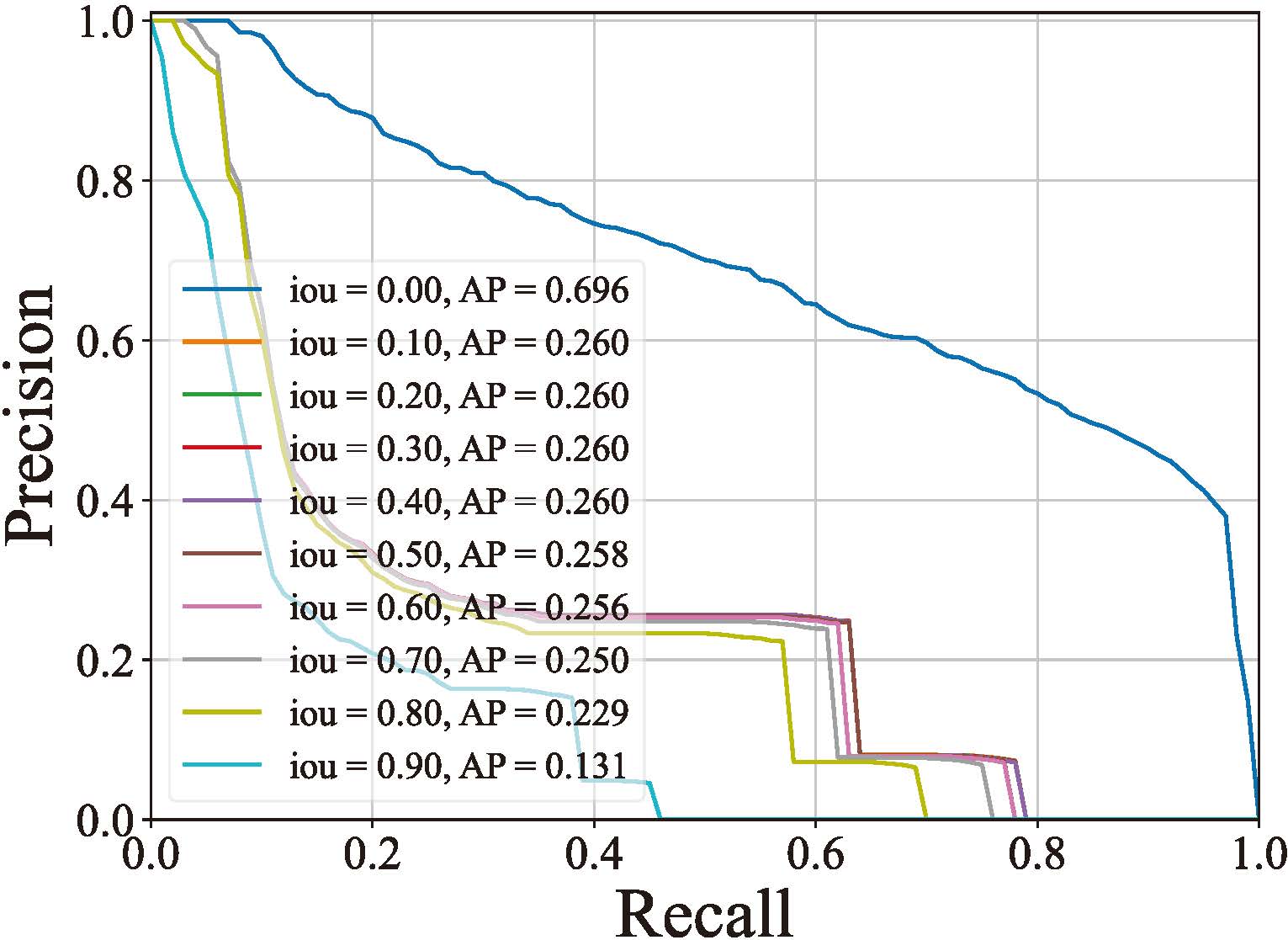}
}
\caption{Figure (a) presents the PR curve of detection results from clear, unprocessed image, while Figure (b) displays the detection results from noisy, unprocessed images. As shown in these figures, noise and blur will introduce very strong effects to detection results and we need to design some algorithms to reduce these effects. In these figures, we can find that if we set smaller IOU, the AP will be larger, which means we can obtain higher recall rate and precision rate at the cost of lower position regression accuracy.}
\label{Fig8}
\end{figure}

In this subsection, we will show the performance of the detection algorithm with the CSST simulation data. Since there are no data pre-processing steps used in this subsection, we use the detection results as the baseline. Two sets of simulated data are used in this subsection. The CSST simulation data with no PSF variations, read out noise or back ground noise are considered as clear data and CSST simulation data with realistic PSF variations and different noises are considered as noisy data. Two neural networks are trained with clean data and noisy data separately through the same procedure. The results are shown in figure  \ref{Fig8}.\\

We designate RT3 to denote clear data and RT4 for noisy data. As illustrated in Figure \ref{Fig8}(\subref{subfig:a}) and Table \ref{table1}, setting $IOU$ to 0 yields a precision rate of approximately 30\%, while the recall rate is 90\%. This indicates that the model can effectively detect nearly all strong lensing systems, albeit with only 30\% of these detections being genuine strong lensing systems. However, when $IOU$ is set to 0.5, as shown in the RT3 column in Table \ref{table2}, there is a noticeable decrease in the recall rate to 70\%, suggesting the model could identify roughly 70\% of all strong lensing systems. In Figure \ref{Fig8}(\subref{subfig:b}) and Table \ref{table1}, the influence of noise becomes apparent, resulting in a detection accuracy of only 69.6\% when $IOU$ is set at 0. With a recall rate of 80\%, the precision rate drops to merely 53.3\%, indicating that only about half of the detected results are strong lensing systems, while the majority are false positives. This low precision rate would necessitate significant human intervention in practical applications. Moreover, with increasing $IOU$ values, as exemplified by the data in the RT4 column of Table \ref{table2}, the accuracy rate sharply decreases, reaching 25.5\% when $IOU$ is set to 0.5. Based on the outcomes presented above, it is apparent that while our methods can yield some results, there are still certain aspects that need improvement. When employing a low $IOU$ threshold, it is expected that numerous detection results would exhibit relatively low precision rates, demanding extensive subsequent human interventions. Consequently, there is a need to further refine our approach and develop complementary methods that can enhance the efficiency of detection.\\

\subsection{Performance Evaluation of the Detection Pipeline}
\label{subsect:PipePer}
As discussed earlier, in order to obtain scientific outcomes with acceptable costs from observation data collected by the CSST, there is an urgent need to enhance the detection capability of our algorithm. Given the impact of varying levels of noise and variable PSFs on images, one strategy involves leveraging image restoration algorithms such as the PSF-Net proposed by \citet{lv2022general,jia2024image}. This involves training the PSF-Net with CSST PSFs and subsequently employing it to enhance the quality of observation images. Considering that many strong lensing systems exhibit relatively low signal-to-noise ratios, we advocate the utilization of the gray scale transformation algorithm outlined in \citet{lupton2004preparing}. This approach renders dim celestial objects easier to be detected. Our proposed strategy involves training the swin transformer-based detection algorithm with a mix of CSST simulation data containing both noisy and clear images. Subsequently, we deploy this trained swin transformer to process images at various stages within the data processing pipeline, thereby providing clearer insights into the necessity of developing a robust data detection pipeline.\\

\begin{figure}
\centering
\subcaptionbox{\label{subfig:a}}{
\includegraphics[width=0.45\textwidth]{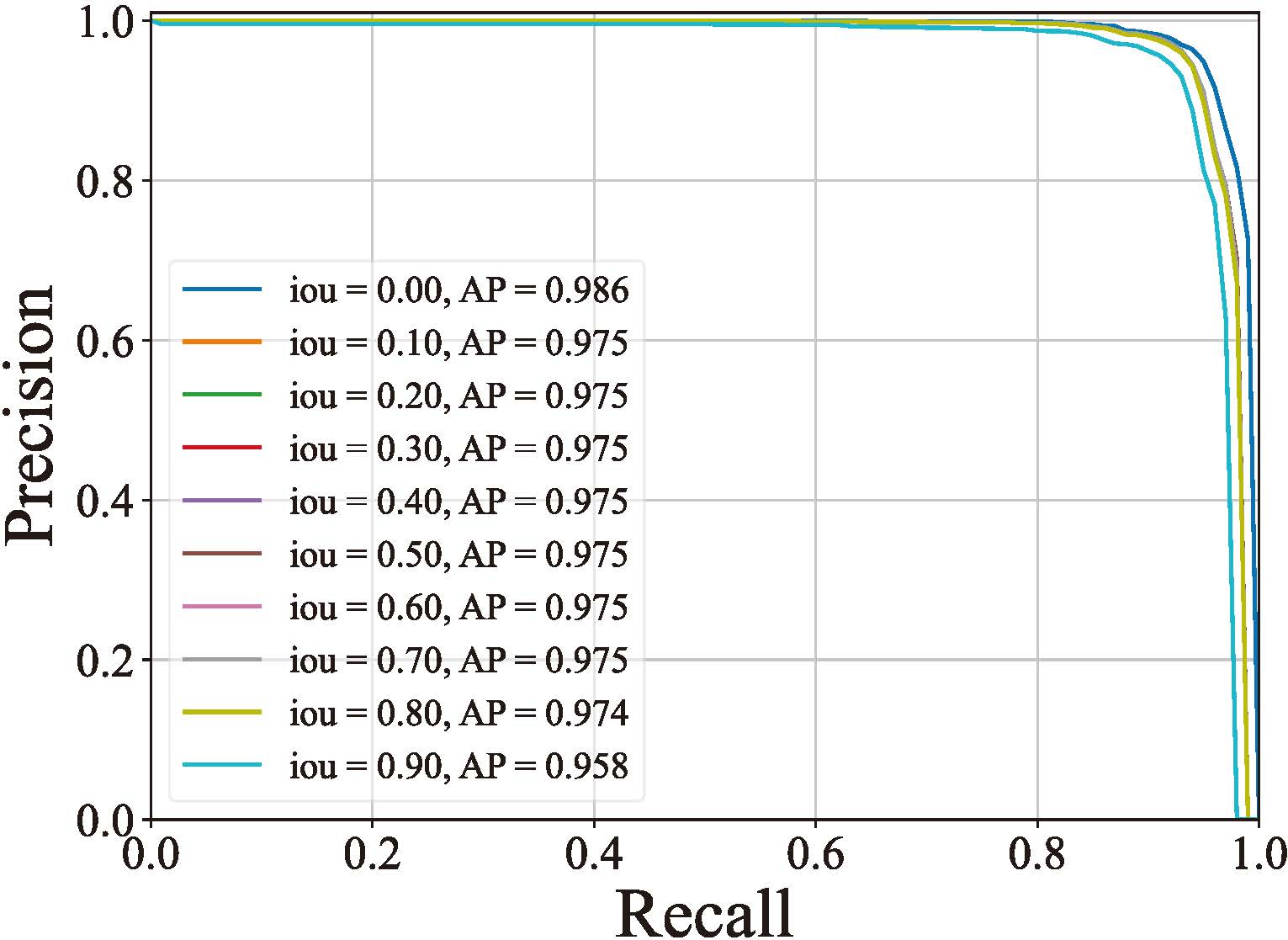}
}
\subcaptionbox{\label{subfig:b}}{
\includegraphics[width=0.45\textwidth]{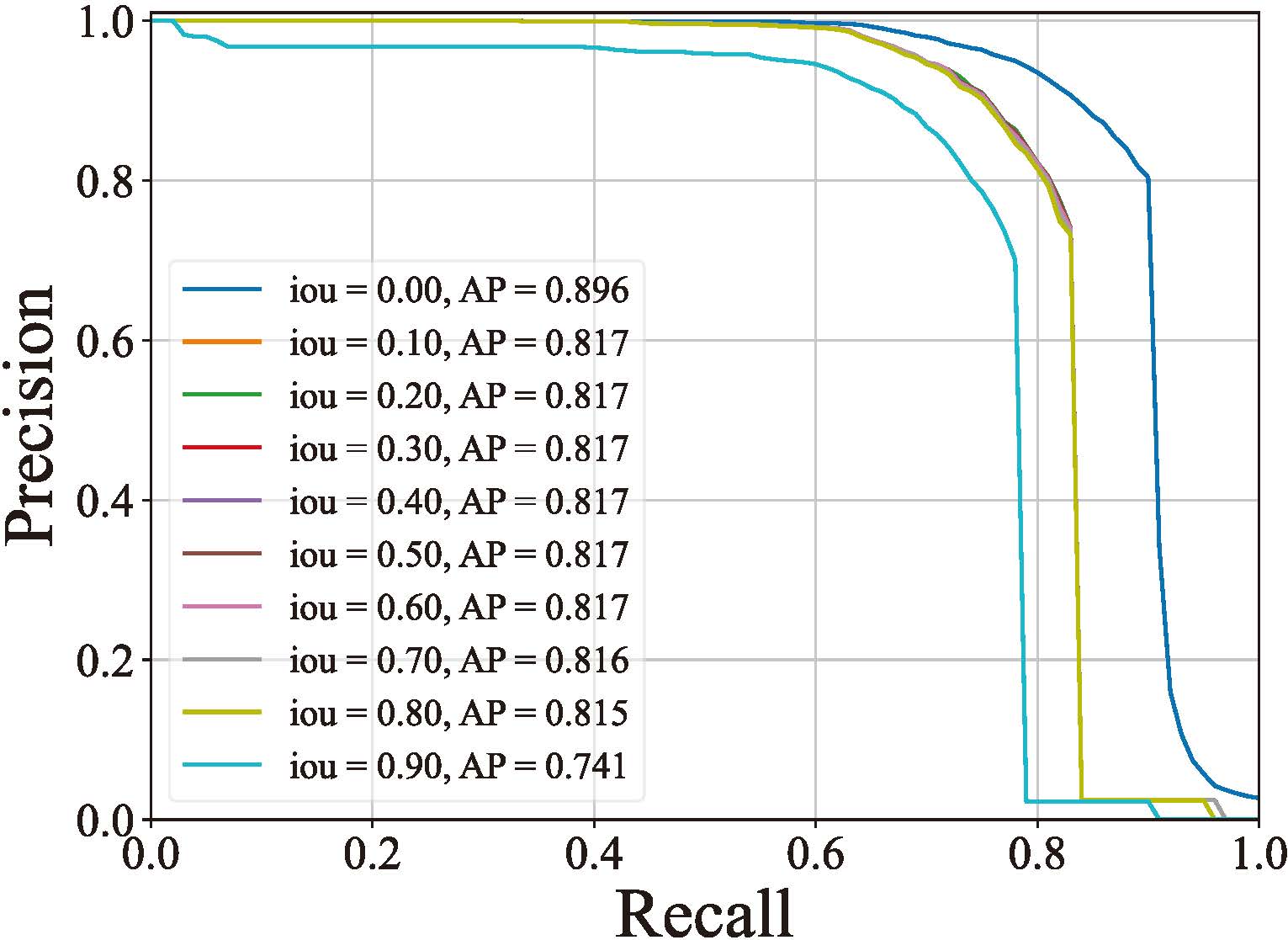}
}
\subcaptionbox{\label{subfig:c}}{
\includegraphics[width=0.45\textwidth]{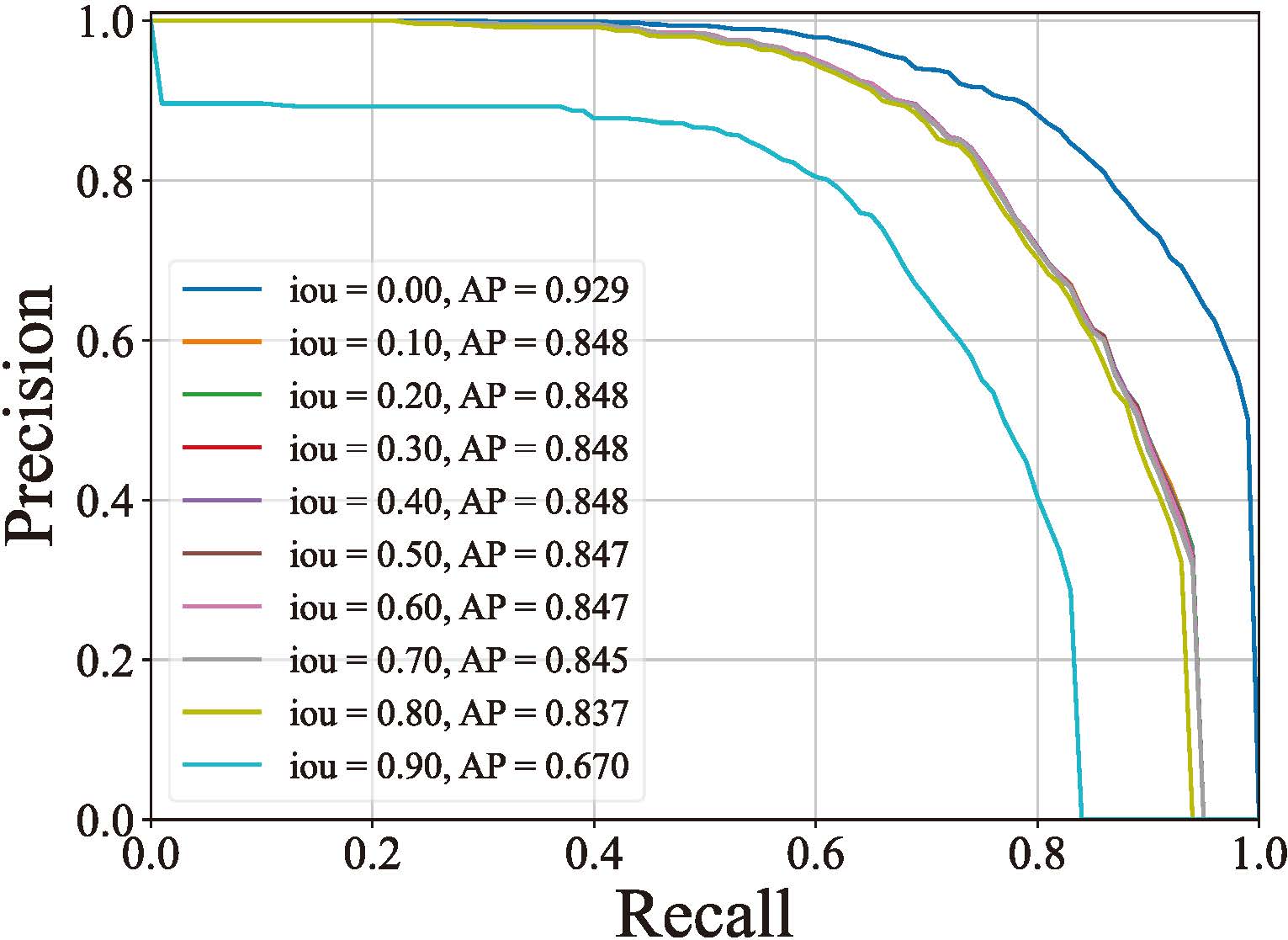}
}
\caption{In Figure (a), the detection results of the clear image after gray scale transformation are depicted. Figure (b) illustrates the PR (Precision-Recall) curve of the noisy image after gray scale transformation, and Figure (c) displays the PR curve of noisy images after gray scale transformation and image restoration. Comparative to the PR curve shown in Figure \ref{Fig8}, the image pre-processing method enhances the detection capabilities of our algorithm and mitigates the effects caused by noise and blur.}
\label{Fig9}
\end{figure}

We test the performance of our algorithm using three sets of simulated images, and the results are depicted in Figure~\ref{Fig9}. In Figure~\ref{Fig9}(\subref{subfig:a}), it is evident that clear images after gray scale transformation demonstrate outstanding performance, achieving a precision of 99\% within a recall range up to 80\%. This highlights the capability of the model to almost flawlessly identify nearly all strong lensing systems with exceptional accuracy. A slight decline is observed only at an IOU of 0.9, attributable to the stringent precision requirement for localization, leading to potential undetected strong lensing systems. In comparison to unprocessed clear images, the detection accuracy rises to 0.975 at an IOU of 0.5, representing a substantial 66\% enhancement. The initial test indicates that gray scale transformation is a crucial step in data preprocessing. We utilize the asinh transformation to process noisy images, followed by employing the trained Swin Transformer to detect strong lensing systems in these images. The detection results are presented in Figure~\ref{Fig9}(\subref{subfig:b}). We use RT3 as clear data, RT4 as noisy data, RT3+asinh and RT4+asinh as asinh gray scale transformation, and RT4+restore+asinh as images after restoration and asinh transformation. In both Table~\ref{table1} and Table~\ref{table2}, when comparing RT3+asinh and RT4+asinh, regardless of whether the IOU is 0 or 0.5, the mAP of noisy images is smaller than that of clear images. Moreover, as the recall rate ranges from 0.8 to 1.0, the precision decreases significantly, resulting in the loss of many strong lensing events. Therefore, we have decided to perform image restoration on the noisy images.\\

\begin{table*}
\centering
\caption{We chose an IOU of 0.0 to generate a table and assess the performance of the model on various datasets. In this table, RT3 denotes clear data, RT4 represents noisy data, RT3+asinh and RT4+asinh indicate data subjected to asinh grayscale transformation, and RT4+restore+asinh is used for noisy images that underwent both image restoration and asinh transformation.}
\begin{tabular}{c|ccccccc} 
\hline
\diagbox{Recall}{Precision}{Data} & {RT3} & {RT4}& {RT3+asinh} & {RT4+asinh}  &{RT4+restore+asinh} \\
\hline 
\multicolumn{1}{c|}{0.0}    & 1.000 & 1.000 & 1.000 & 1.000 & 1.000  \\
\multicolumn{1}{c|}{0.1}    & 1.000 & 0.981 & 1.000 & 1.000 & 1.000  \\
\multicolumn{1}{c|}{0.2}    & 0.998 & 0.879 & 1.000 & 1.000 & 1.000  \\
\multicolumn{1}{c|}{0.3}    & 0.994 & 0.810 & 1.000 & 1.000 & 1.000  \\
\multicolumn{1}{c|}{0.4}    & 0.977 & 0.746 & 1.000 & 0.999 & 0.996  \\
\multicolumn{1}{c|}{0.5}    & 0.940 & 0.701 & 1.000 & 0.999 & 0.992  \\
\multicolumn{1}{c|}{0.6}    & 0.891 & 0.645 & 1.000 & 0.997 & 0.976  \\
\multicolumn{1}{c|}{0.7}    & 0.816 & 0.597 & 0.999 & 0.980 & 0.933  \\
\multicolumn{1}{c|}{0.8}    & 0.574 & 0.533 & 0.998 & 0.935 & 0.865  \\
\multicolumn{1}{c|}{0.9}    & 0.436 & 0.465 & 0.984 & 0.804 & 0.728  \\
\multicolumn{1}{c|}{1.0}    & 0.000 & 0.000 & 0.000 & 0.027 & 0.322  \\
\bottomrule
\end{tabular}
\label{table1}
\end{table*}

\begin{table*}
\centering
\caption{We chose an IOU of 0.5 to generate a table and assess the performance of the model on various datasets. In this table, RT3 denotes clear data, RT4 represents noisy data, RT3+asinh and RT4+asinh indicate data subjected to asinh grayscale transformation, and RT4+restore+asinh is used for noisy images that underwent both image restoration and asinh transformation.}
\begin{tabular}{c|ccccccc} 
\hline
\diagbox{Recall}{Precision}{Data} & {RT3} & {RT4}& {RT3+asinh} & {RT4+asinh} & {RT4+restore+asinh} \\
\hline 
\multicolumn{1}{c|}{0.0}    & 1.000 & 1.000 & 1.000 & 1.000 & 1.000   \\
\multicolumn{1}{c|}{0.1}    & 0.995 & 0.639 & 1.000 & 1.000 & 1.000  \\
\multicolumn{1}{c|}{0.2}    & 0.992 & 0.334 & 1.000 & 1.000 & 0.998  \\
\multicolumn{1}{c|}{0.3}    & 0.974 & 0.270 & 1.000 & 1.000 & 0.997   \\
\multicolumn{1}{c|}{0.4}    & 0.913 & 0.255 & 1.000 & 0.999 & 0.990   \\
\multicolumn{1}{c|}{0.5}    & 0.816 & 0.255 & 1.000 & 0.997 & 0.978  \\
\multicolumn{1}{c|}{0.6}    & 0.642 & 0.251 & 0.999 & 0.992 & 0.949     \\
\multicolumn{1}{c|}{0.7}    & 0.226 & 0.080 & 0.998 & 0.948 & 0.862   \\
\multicolumn{1}{c|}{0.8}    & 0.000 & 0.000 & 0.997 & 0.822 & 0.706   \\
\multicolumn{1}{c|}{0.9}    & 0.000 & 0.000 & 0.981 & 0.024 & 0.477  \\
\multicolumn{1}{c|}{1.0}    & 0.000 & 0.000 & 0.000 & 0.000 & 0.000  \\
\bottomrule
\end{tabular}
\label{table2}
\end{table*}

Finally, we combine both the image restoration and gray scale transformation algorithms to process noisy images and assess the performance of the detection pipeline. The results serve as a reference for anticipating potential scientific outputs of the CSST in the detection of strong lensing systems, as depicted in Figure~\ref{Fig9}(\subref{subfig:c}). With an IOU of 0, the detection accuracy is 92.9\%. Although the improvement in mAP might not be dramatic, there is a noteworthy increase in the recall rate when the IOU ranges between 0.2 and 1.0. This is crucial since the initial maximum recall rate is 90\%, indicating that at least 10\% of strong lensing systems may go undetected regardless of parameter adjustments. However, following image restoration, the recall rate reaches 100\%, significantly reducing the number of undetectable targets. Hence, the image restoration algorithm proves indispensable and justifies its integration into the detection pipeline. The detection results demonstrate substantial improvement with the pipeline. In practical scenarios, configuring the IOU threshold to 0.5 yields a recall rate of 0.6 and a precision rate of 0.949. Considering the vast number of galaxies and potential strong lensing systems observable using the CSST, we have the potential to identify strong lensing candidates. However, within this large pool, thorough verification by human experts is necessary, posing a significant challenge. To address this, we are exploring collaboration with citizen science projects to assist in obtaining the final validated detection outcomes\footnote{\url{https://nadc.china-vo.org/lensfinder}}, which is inspired by the GalaxyZoo projects \citep{fortson2012galaxy}.\\

\subsection{Visual Analysis of Detection Results}
\label{subsect:AnalyseResults}
\begin{figure}
\centering
\includegraphics[width=0.3\textwidth]{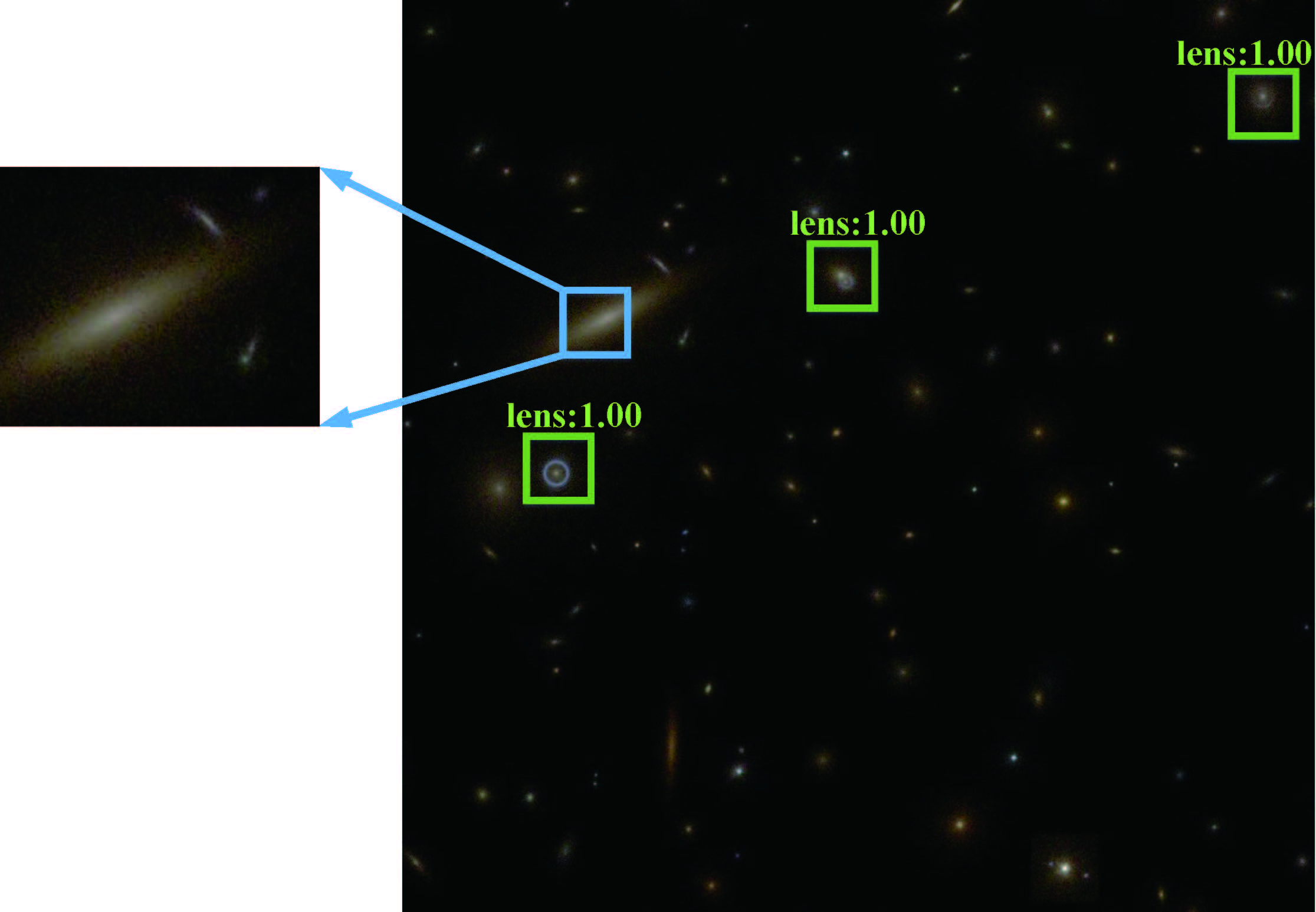}
\caption{The utilization of a fixed-size bounding box may affect detection outcomes, especially when dealing with large-sized strong lensing systems. Illustrated in this figure, the blue box denotes a specific strong lensing object, but the bounding box only partially encloses it. As a result, the IOU score becomes excessively low, hindering the accurate detection of this particular strong lensing event.}
\label{Fig10}
\end{figure}

It is crucial to examine the outcomes of the detection process and assess potential issues and risks associated with deploying the detection algorithm. We have evaluated four distinct scenarios and provide detailed discussions for each of them below:\\
1. Impact of Strong Lensing Systems with Large Size: Strong Lensing systems often have different sizes. In the label creation step, we standardize the bounding boxes to dimensions of $50 \times 50$ pixels. This uniform size facilitates efficient detection by the neural network while mitigating the influence of background and noise. This approach generally proves effective for most strong lensing systems. However, for certain instances with larger spatial extents, their structures surpass these bounding boxes. This over-extension disrupts their structural features and engenders information loss. Consequently, our neural network encounters difficulty in accurately pinpointing these extended gravitational lensing systems. As shown in the blue box in the top-left corner of figure \ref{Fig10}, the blue boxes represent the strong lensing systems present in the labels, while the green boxes indicate the correctly detected strong lensing systems. The strong lensing system on the upper left is notably large. The bounding box with size of $50 \times 50$ pixel leads to a misplacement of the detection position at the center of the bounding box for this extended object. This misalignment results in flawed feature extraction by the model, yielding imprecise detection results. Normally the strong lensing systems larger than $50 \times 50$ pixels would hard to be located accurately.\\

\begin{figure}
\centering
\subcaptionbox{\label{subfig:a}}
    {
    \includegraphics[width=0.3\textwidth]{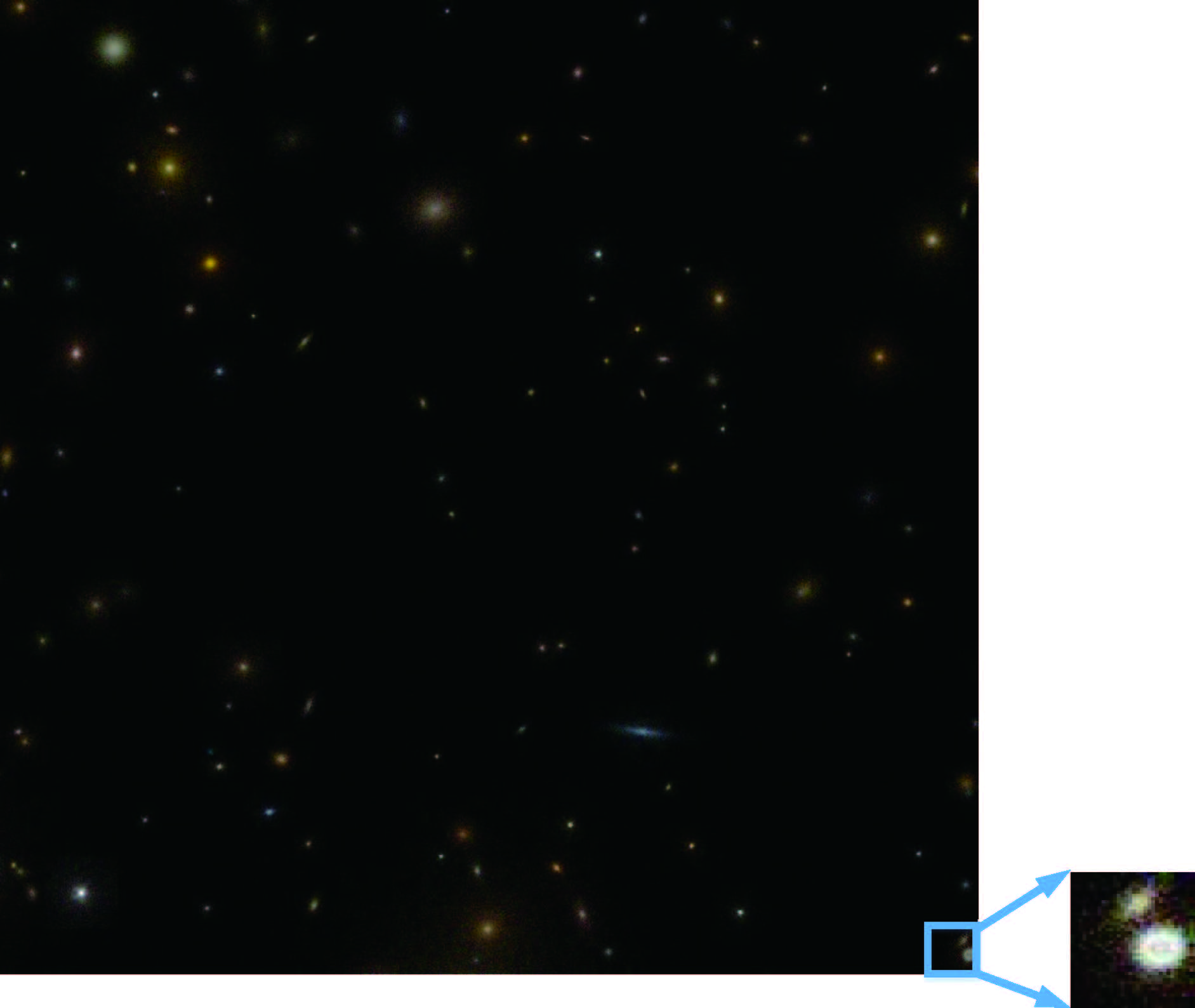}}
\subcaptionbox{\label{subfig:b}}
    {
    \includegraphics[width=0.32\textwidth]{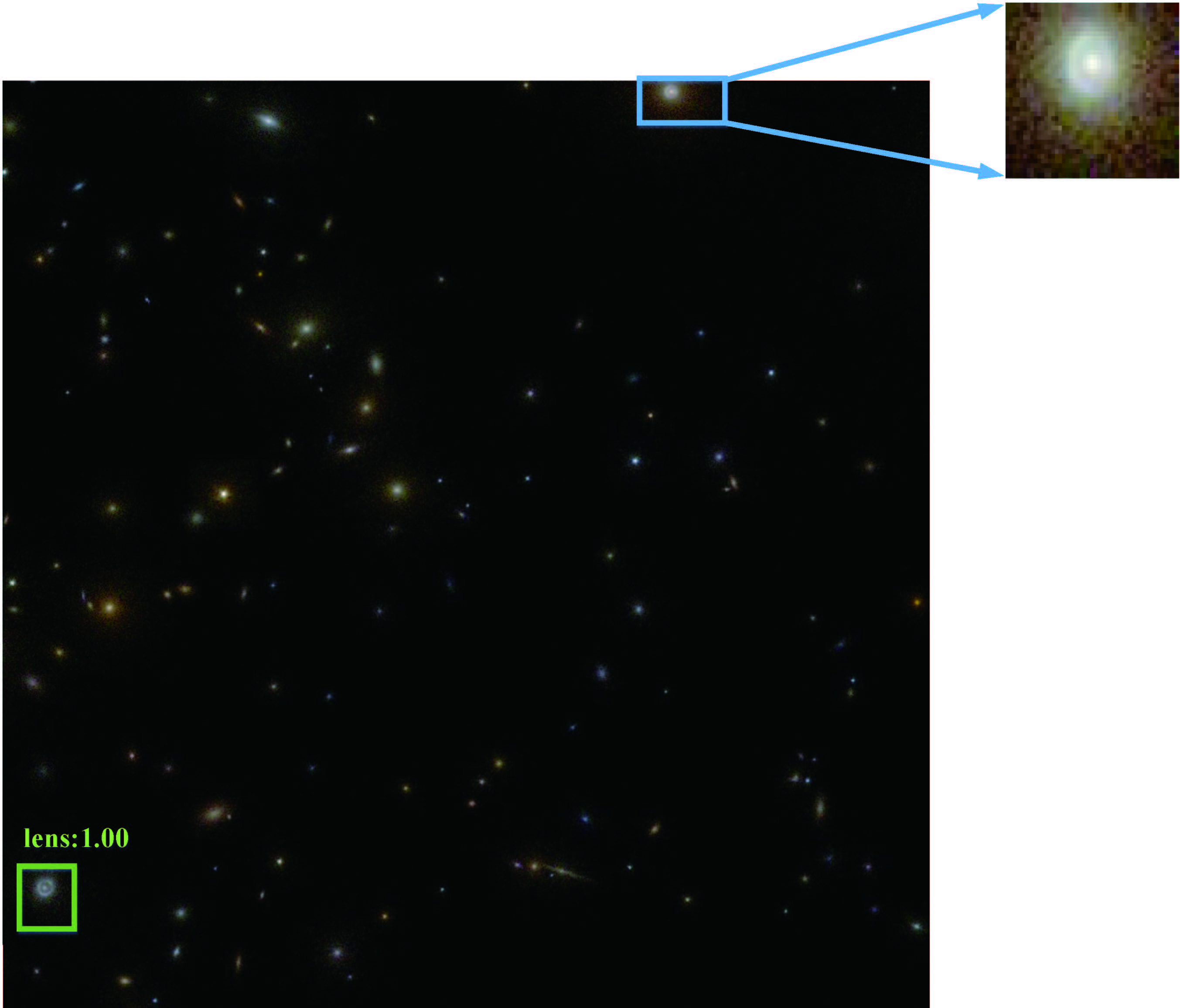}}
\caption{The strong lensing system highlighted within the blue box in Figure (a) exhibits dimensions of $16\times 17$ pixels and a signal-to-noise ratio (SNR) of 15.88. However, due to its small size, our algorithm was unable to detect it. The strong lensing system indicated by the blue box in Figure (b) has dimensions of $25\times 25$ pixels and an SNR of 10.42. Despite its larger size, its relatively low SNR prevented our algorithm from detecting it as well.}
\label{Fig11}
\end{figure}

2. Impact of Strong Lensing Systems with Small Size or Low Signal-to-Noise Ratio: Distinctive curved structures and the presence of multiple images are features of strong lensing systems. However, the detection of these strong lensing systems becomes particularly challenging when they are exceptionally small or exhibit low signal-to-noise ratios. As illustrated in the left panel of Figure \ref{Fig11}(\subref{subfig:a}), the lens structure in the bottom right corner of the image is extremely small, making it prone to identification as background or other objects. In the right panel of Figure \ref{Fig11}(\subref{subfig:b}), the strong lensing system is heavily impacted by background noise, making it difficult for the model to accurately recognize the objects. Both small- and low-signal-to-noise ratio strong lensing systems could be categorized as targets with a low signal-to-noise ratio, defined as $SNR = TotalFlux\_StrongLensing/Variance\_Background$ (the ratio of total photons within strong lensing systems to the variance of background noise). Typically, identifying strong lensing systems in CSST observation data with a signal-to-noise ratio less than 16 is considerably challenging, with 0.36\% of them being effectively undetected.\\

\begin{figure}
\centering
\subcaptionbox{\label{subfig:a}}{
\includegraphics[width=0.3\textwidth]{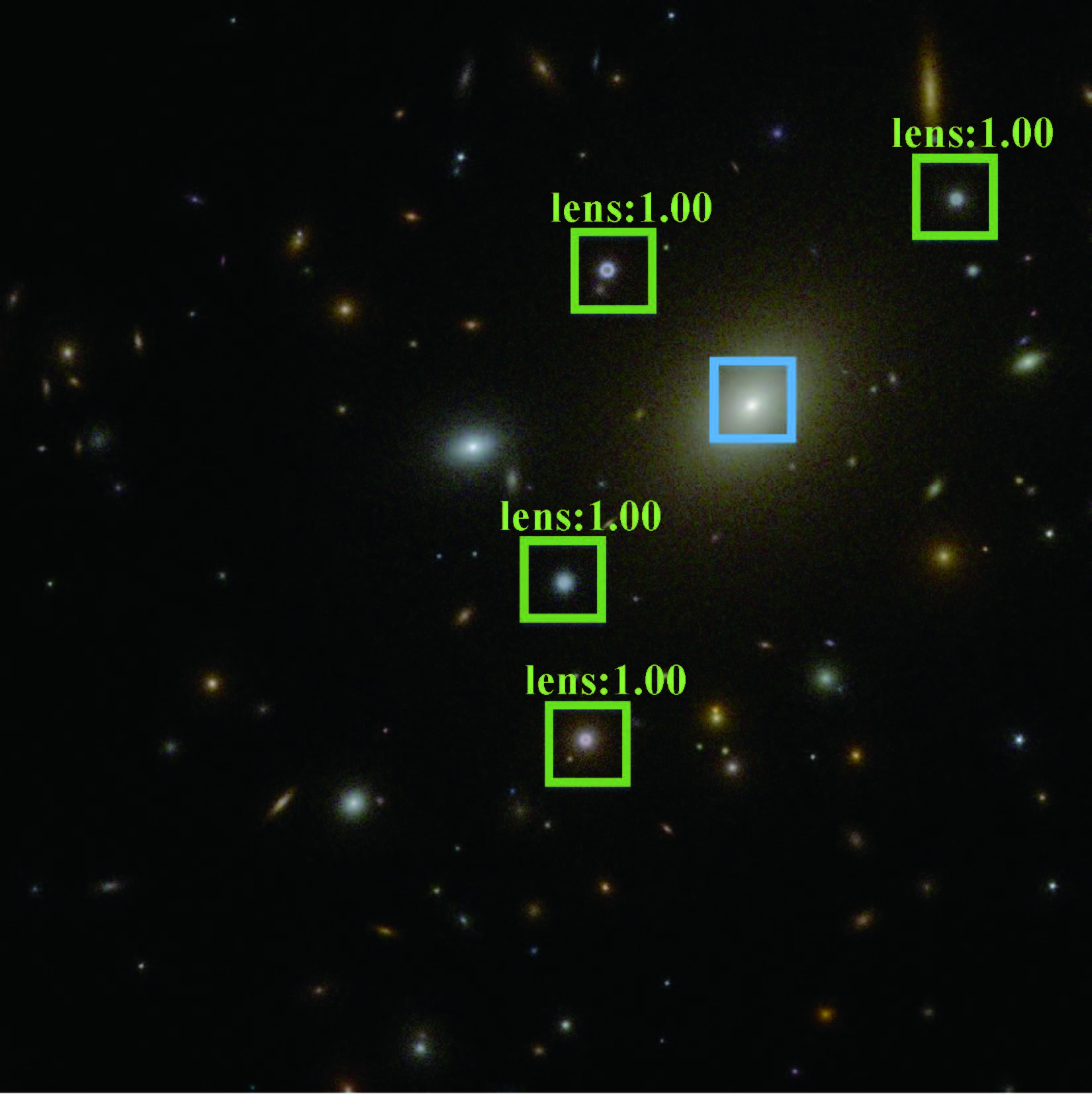}
}
\subcaptionbox{\label{subfig:b}}{
\includegraphics[width=0.3\textwidth]{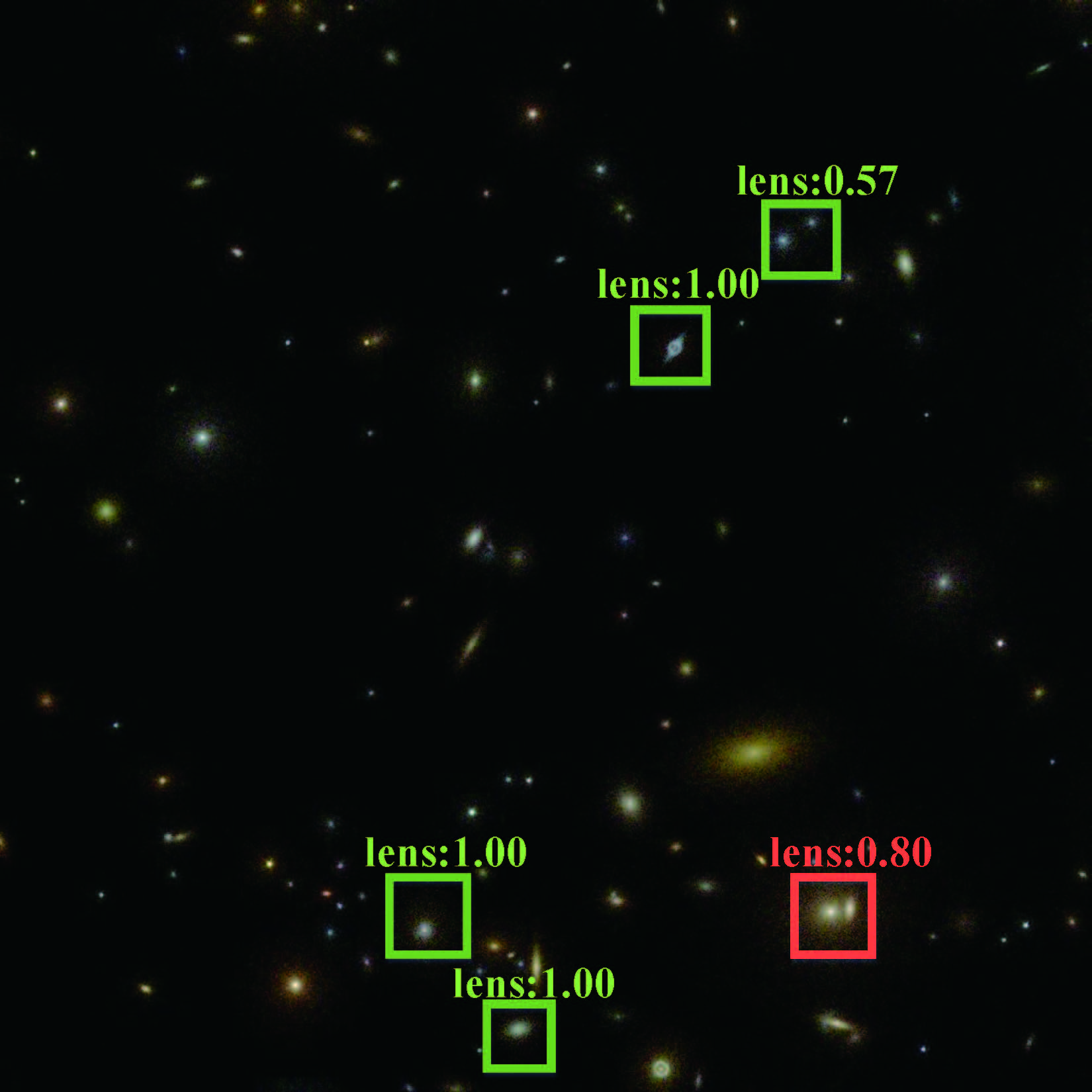}
}
\caption{The presence of bright central galaxy light can pose challenges in detecting strong lensing, as illustrated by the blue bounding box in Figure (a). Figure (b) depicts several bright stars outlined with green bounding boxes, which were detected as strong lensing systems. This could be attributed to their visual similarity to actual strong lensing systems.}
\label{Fig12}
\end{figure}

3. Impact of Bright Sources: Bright central galaxies and bright stars will affect the detection efficiency. Light from the bright central galaxies will make light from the strong lensing systems hard to recognize in multiple color space defined by filters of the CSST, thus these targets will not be detected in observation images. As shown in  figure \ref{Fig12}(\subref{subfig:a}), there is a strong lensing system behind the galaxy, which are hard to be detected by our algorithm. Meanwhile, several bright stars with similar colour are very likely to be detected as strong lensing systems, which is probably caused by the similaritiy between them and strong lensing systems. As shown in the right panel of figure \ref{Fig12}, the system in red colour stands for a false detection. Detection of strong lensing system regardless of bright central galaxies is a challenge for human beings. We can find that although our algorithm partly solves this problem through looking into the image with more colours, there are still gaps to fill. We need to further consider channel attention to design some novel algorithms to solve this problem. However it should be mentioned that the problems brought by bright stars may not be a problem, since we can mask these targets before we start to detect strong lensing systems.\\

\section{Detection of Strong Lensing Systems from Real observation Data}
\label{sect:realapp}
We evaluate the effectiveness of our pipeline using real observation data. Our testing involves processing images obtained from the DESI Legacy Imaging Surveys and media images from Euclid Early Release Observations using our pipeline. The images from the DESI Legacy Imaging Surveys contain numerous known strong lensing systems, providing an opportunity to assess the performance of our method. Despite the lower gray scale levels in the media images from Euclid, which is a space telescope with similar spatial resolution, we can still make a preliminary estimation of the scientific capabilities of the CSST in detecting strong lensing systems by using Euclid images as a reference.\\

\subsection{Detection of Strong Lensing Systems from DESI Legacy Imaging Surveys}
\label{sect:realapp}
We use the following steps to test the performance of our pipeline with images from the DESI Legacy Imaging Surveys:
\begin{itemize}
\item We have collected images from the DESI Legacy Imaging Surveys DR9 \citep{huang2020finding} that contain 5060 known strong lensing systems. Due to hardware limitations, we have resized these larger images to dimensions of $1000 \times 1000$ pixels, resulting in a dataset of 12667 images. In this dataset, we also split the data into the training set and the validation set in a 7:3 ratio.
\item Given the random distribution of strong lensing systems in actual data, our model is designed to identify such systems regardless of their placement. This means that no matter where the strong lensing systems is situated within the image, our model should be adept at detection. Consequently, when generating training data, we deliberately shuffle the positions of strong lensing systems to simulate different scenarios. With the above steps, we have obtained the training data.
\item We begin by choosing the detection neural network that has been trained using CSST simulation data as the initial weights. Subsequently, we employ the specified training data discussed earlier to fine-tune this neural network. With the learning rate set at 0.001, we have train the neural network over 100 epochs.
\item After training, the neural network becomes capable of identifying strong lensing systems from images obtained from the DESI Legacy Imaging Surveys. To assess its performance, we utilize the entire set of 5060 images to evaluate the effectiveness of our pipeline, aiming to ascertain its ability to detect recognized strong lensing systems and also discover previously known strong lensing systems.
\end{itemize}

\subsubsection{Detection Pipeline Test with Known Strong Lensing Systems}
\label{subsect:stronglensingknown}
\begin{figure}
\centering
\subcaptionbox{}{
\includegraphics[width=0.25\textwidth]{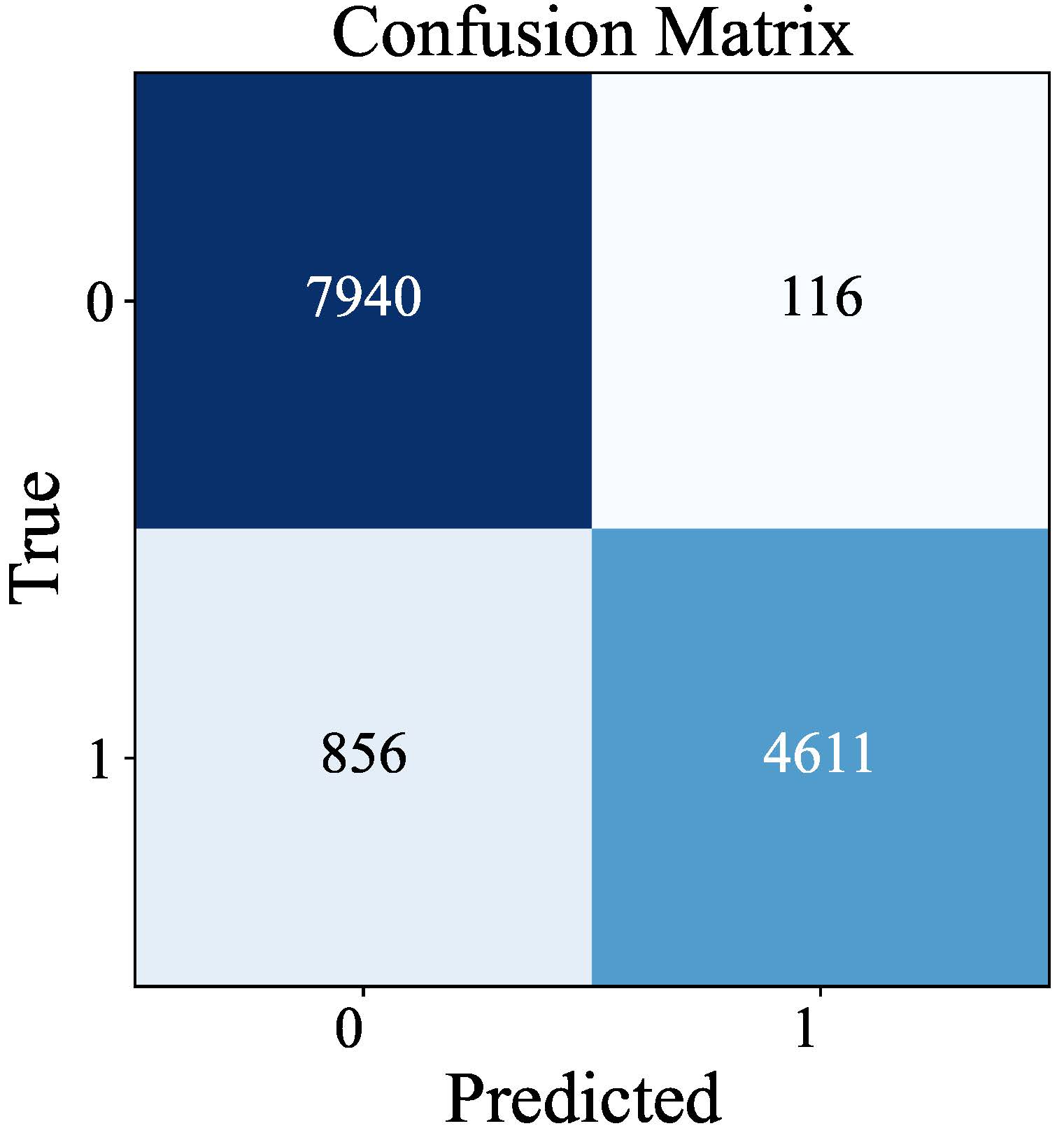}
}
\subcaptionbox{}{
\includegraphics[width=0.25\textwidth]{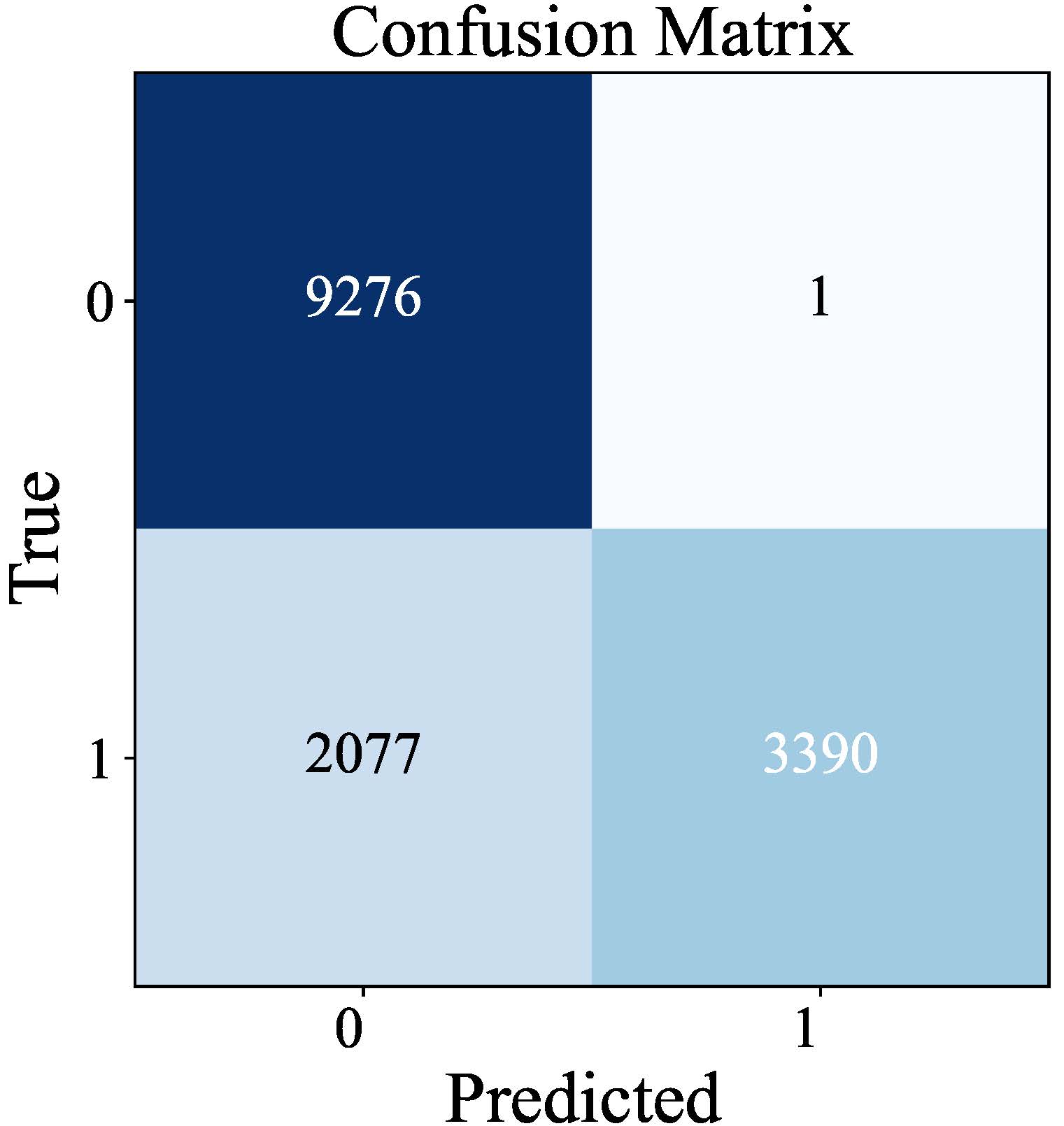}
}
\subcaptionbox{}{
\includegraphics[width=0.25\textwidth]{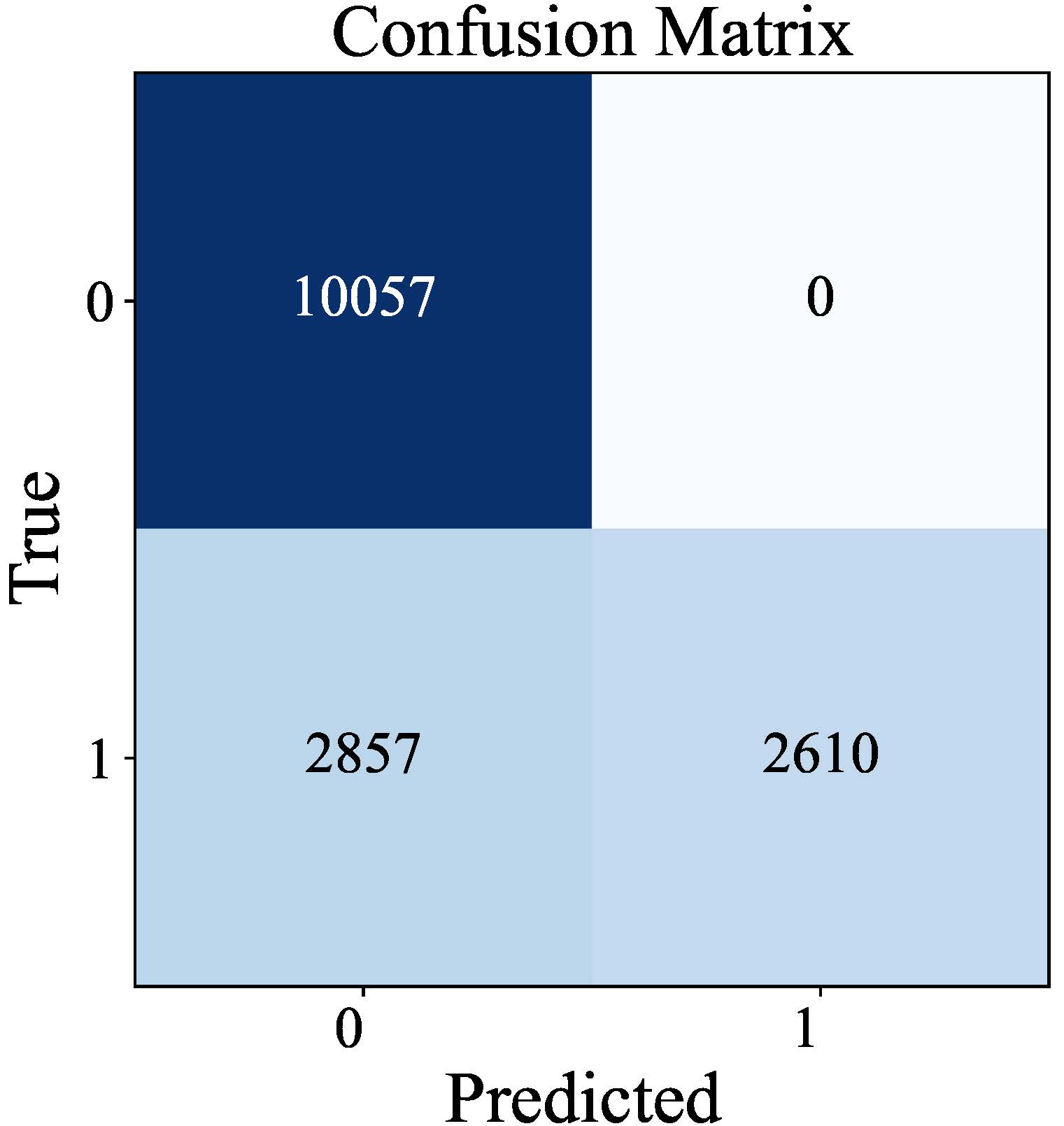}
}
\caption{Figures (a), (b), and (c) depict the confusion matrix of our detection algorithm using confidence score threshold values of 0, 0.5, and 0.9 on the simulated data. As demonstrated in these figures, our algorithm performs well in detecting strong lensing systems, especially considering the significant influence of bright central galaxies and low signal-to-noise ratios on many of these systems. Additionally, it is important to highlight that we can achieve a satisfactory balance of precision and recall rates for further analysis by selecting a confidence score threshold of 0.5.}
\label{Fig13}
\end{figure}

For detection of known strong lensing systems, we have set the confusion matrix with confidence score threshold of 0, 0.5 and 0.9 in figure \ref{Fig13}. When the threshold is set to 0, the recall rate of the model is 84.3\%, indicating that the model is able to capture a certain proportion of true strong lensing systems. At this threshold, the precision is 97.54\%, meaning that 97.54\% of the samples detected by the model are indeed true strong lensing systems, with only about 2.5\% being false positives. As the threshold increases, the number of false positives gradually decreases, indicating a continuous improvement in the precision of the model. Particularly, at a threshold of 0.5, there are only one false positives, and the precision reaches 99.97\%. This demonstrates that the detection ability and accuracy of the model significantly improve at higher thresholds. At a threshold of 0.9, there are no false positives, resulting in a precision of 100\%. This further confirms that our pipeline has good performance in processing real observation data.\\

It is worth noting that the low false positive rate implies that when applying the model to large-scale sky survey data for detecting galaxy-scale lensings, there will not be a significant number of falsely reported strong lensing cases. This is a crucial property as reducing false alarms can save valuable time and resources in research, while enhancing the overall credibility of the study. These results provide crucial metrics for evaluating the performance of the model in identifying strong lensing systems in real data. It is evident that by adjusting the threshold, we can control the accuracy and recall rate of the detection results. Depending on specific requirements, we can choose a higher threshold (such as 0.9) to reduce false positives or lower the threshold to improve the recall rate.\\

\subsubsection{Detection Pipeline Test For Discovery of New Strong Lensing Systems}
\label{subsect:stronglensingunknown}
\begin{figure*}
\centering
\includegraphics[width=0.9\textwidth]{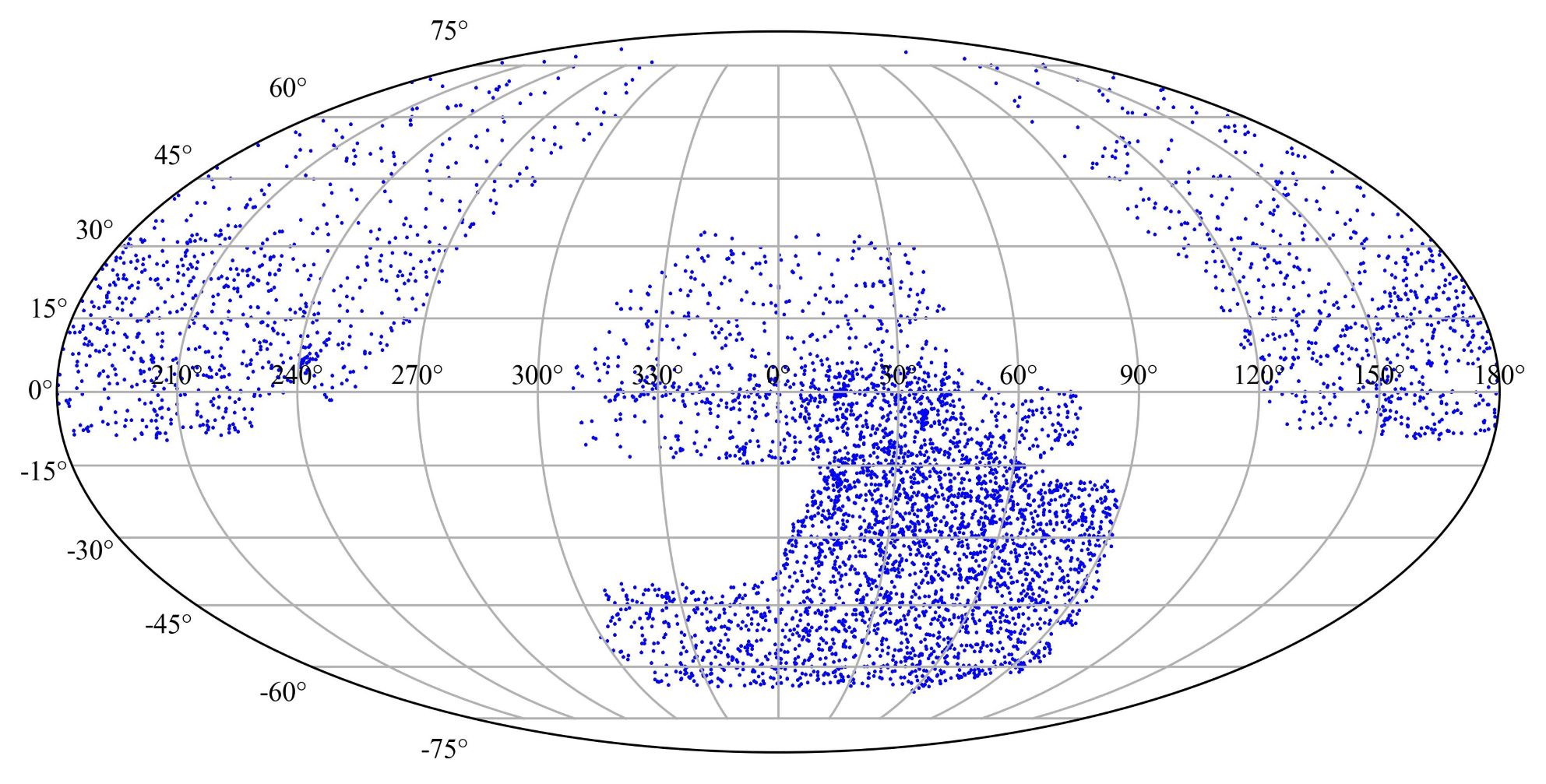}
\caption{The images selected for detecting strong lensing systems follow the distribution outlined in \citet{huang2020finding}. It is important to highlight that only a small subset of these images is utilized in this paper. Further elaboration on this issue will be provided in our future papers.}
\label{Fig14}
\end{figure*}

While testing the performance of our algorithm on data from the DESI Legacy Imaging Survey DR9, we have identified several new strong lensing candidates. With a detection threshold set at 0.5, we have applied the pipeline to process the images illustrated in Figure~\ref{Fig14}. The dataset also comprises 5060 images, each with dimensions of $1000\times 1000$ pixels, covering an area of 241.207 $deg^2$. Following the procedures outlined in Section \ref{sect:data} and Section \ref{sect:switrans}, these images are processed. The PSF-Net was trained using CSST images blurred by the Moffat model with variable FWHM values ranging from 2.0 to 8.0 pixels. After preprocessing and gray scale transformation, the neural network discussed in Section \ref{subsect:stronglensingknown} is utilized for further image processing. The comprehensive task of processing all these images takes 72 hours and results in the identification of additional candidates beyond the known 5060 instances of strong gravitational lensings. In Appendix \ref{appedix}, we provide the celestial coordinates of some these newly identified strong lensing candidates. These findings indicate that the majority of them exhibit arc-like structures or multiple images, emphasizing the need for further follow-up observations. To facilitate such endeavors, we have included a complete table of strong lensing systems in this paper, providing users with a comprehensive resource for undertaking subsequent observations.\\

\subsection{Detection of Strong Lensing Systems from Media Images from Euclid Early Data Release}
\label{sect:realapp2}
Euclid, a spaceborne telescope developed and operated by the European Space Agency, boasts a 1.2-meter aperture and a spatial resolution of 0.2 arcseconds. Given that the Euclid observes across the visible (550 nm) to near-infrared (900 nm) band and shares a spatial resolution similar to that of the CSST, its images can effectively assess the performance of our pipeline. To evaluate our algorithm, we utilize the media images released by Euclid on November 7, 2023. For testing purposes, we exclude the Horsehead Nebula because of its complex background, which could introduce additional errors into the detection algorithm. As these media images have already undergone processing by pipelines developed by Professor Jean-Charles Cuillandre, we directly apply our algorithm to detect strong lensing systems without the need for additional data preprocessing. However, given that the media images are tiff files, which have relatively low gray scale levels and are composed of data from both visible and near-infrared bands, the algorithm may make errors since it was trained on simulated CSST data covering the near-ultraviolet to visible band. We extract stamp images with dimensions of $200 \times 200$ pixels from the original images and set the confidence threshold for detection at 0.5. To ensure a complete detection of strong lensing systems, we also check the results by eyes.\\

In total, our algorithm identified 67 strong lensing system candidates, which we categorize into three groups: strong lensing systems with high quality, strong lensing systems with moderate quality, and false positive results. The Appendix~\ref{appedixb} shows 15 candidates for high quality strong lensing systems, confirming the effectiveness of our method in detecting strong lensing systems. Additionally, there are 22 strong lensing system candidates with moderate quality, indicating that the completeness and purity of our algorithm can be adjusted by varying the confidence level. However, 30 of the detection results are false positives, mostly attributed to spiral galaxies and spikes generated by the diffraction of the spider. Given that the CSST employs a Three-Mirror Anastigmat system, which does not introduce diffraction arms, we anticipate that these false positives will not pose a serious problem. However, the lack of spiral galaxies with fine details remains a challenge for our algorithm. To address this, we plan to perform high-fidelity simulations to obtain data for further training of the neural network.\\

\section{Conclusions}
Traditional methods for searching for strong lensing systems are known for being time-consuming and labor-intensive. As we look ahead to upcoming sky survey projects, such as the CSST, the development of highly efficient detection algorithms becomes paramount. In response to these urgent needs, this study has developed a comprehensive pipeline. The process begins with the generation of mock observation images based on a catalog of strong lensing systems and other celestial objects, which serve as prior information. Subsequently, a neural network based on the Swin-Transformer architecture is trained using either simulated images or real observation images containing verified strong lensing systems. Through training, the neural network becomes adept at identifying strong lensing systems within observation images. To enhance the detection ablity of the algorithm, we introduce additional image preprocessing steps. These steps involve various operations, including grayscale transformation and image restoration, applied to the original astronomical image data.\\

We rigorously assess the performance of the model using a comprehensive set of evaluation metrics. This assessment encompasses simulated data and real observed data, as well as grayscale transformation applied to the data. When the noise in the data had a significant impact, we introduced an image restoration algorithm, combined with gray scale transformation, to enhance the completeness of the detection. When applied to simulated CSST observation data, our method demonstrates a detection accuracy rate of 98.6\% and a recall rate of 99.79\%. Given the remarkable observation capabilities of the CSST, we are optimistic about its potential to identify undiscovered strong lensing systems. Moreover, we extend the application of our pipeline to some images from the Euclid observation data, where it successfully identifies 37 strong lensing system candidates.\\

Moving forward, our strategic roadmap involves refining the data structure to align the proportion of strong lens data more closely with real situations. This crucial adjustment is anticipated to improve the precision of the detection capabilities of the model. In response to false detection results stemming from small lens/source structures and bright stars, we will enhance the data simulation algorithms by involving HST observations and hydrodynamic simulations to generate training sets adhering more rigorously to real observations, which is particularly important for the reliability of applying our framework to space-borne imaging surveys, such as CSST, Euclid, and Roman\footnote{\url{https://roman.gsfc.nasa.gov/}}. Furthermore, the images from the DESI Legacy Survey DR9 that we have been using are limited, and only several images from Euclid telescopes are used. In the future, our plan involves searching for strong lensing systems from the DESI Legacy Survey DR 10, with the goal of identifying an increased number of strong gravitational lensing system candidates. Expanding our strategy, we plan to undertake joint training utilizing real observation data obtained from different telescopes. This concerted approach seeks to increase the accuracy and recall rate. This comprehensive plan not only promises more scientific outcomes for ongoing initiatives such as the CSST main survey or the data from Euclid but also holds the potential to significantly advance our capacity to discover other celestial objects, ranging from nebulae and galaxies to strong lensing systems.\\

\section*{Acknowledgments}
This paper used data obtained with the Dark Energy Camera (DECam), which was constructed by the Dark Energy Survey (DES) collaboration. Funding for the DES Projects has been provided by the U.S. Department of Energy, the U.S. National Science Foundation, the Ministry of Science and Education of Spain, the Science and Technology Facilities Council of the United Kingdom, the Higher Education Funding Council for England, the National Center for Supercomputing Applications at the University of Illinois at Urbana-Champaign, the Kavli Institute of Cosmological Physics at the University of Chicago, the Center for Cosmology and Astro-Particle Physics at the Ohio State University, the Mitchell Institute for Fundamental Physics and Astronomy at Texas A\&M University, Financiadora de Estudos e Projetos, Fundação Carlos Chagas Filho de Amparo à Pesquisa do Estado do Rio de Janeiro, Conselho Nacional de Desenvolvimento Científico e Tecnológico and the Ministério da Ciência, Tecnologia e Inovacão, the Deutsche Forschungsgemeinschaft, and the Collaborating Institutions in the Dark Energy Survey. The Collaborating Institutions are Argonne National Laboratory, the University of California at Santa Cruz, the University of Cambridge, Centro de Investigaciones Enérgeticas, Medioambientales y Tecnológicas-Madrid, the University of Chicago, University College London, the DES-Brazil Consortium, the University of Edinburgh, the Eidgenössische Technische Hochschule (ETH) Zürich, Fermi National Accelerator Laboratory, the University of Illinois at Urbana-Champaign, the Institut de Ciències de l'Espai (IEEC/CSIC), the Institut de Física d'Altes Energies, Lawrence Berkeley National Laboratory, the Ludwig-Maximilians Universität München and the associated Excellence Cluster Universe, the University of Michigan, the National Optical Astronomy Observatory, the University of Nottingham, the Ohio State University, the OzDES Membership Consortium the University of Pennsylvania, the University of Portsmouth, SLAC National Accelerator Laboratory, Stanford University, the University of Sussex, and Texas A\&M University.\\

This paper used public media images from Euclid. Euclid is a European mission, built and operated by ESA, with contributions from NASA. The Euclid Consortium – consisting of more than 2000 scientists from 300 institutes in 13 European countries, the US, Canada and Japan – is responsible for providing the scientific instruments and scientific data analysis. ESA selected Thales Alenia Space as prime contractor for the construction of the satellite and its service module, with Airbus Defence and Space chosen to develop the payload module, including the telescope. NASA provided the detectors of the Near-Infrared Spectrometer and Photometer, NISP. Euclid is a medium-class mission in ESA’s Cosmic Vision Programme.\\

This work is supported by the National Natural Science Foundation of China (NFSC) with funding numbers 12173027, 12303105, 12173062, 12120101003, 12373010, and National Key Research and Development Program with funding number of 2023YFF0725300. We acknowledge the science research grants from the China Manned Space Project with NO. CMS-CSST-2021-A01, CMS-CSST-2021-A02 and science research grants from the Square Kilometre Array (SKA) Project with NO. 2020SKA0110102.\\

\section*{Data Availability}
Data resources are supported by China National Astronomical Data Center (NADC) and Chinese Virtual Observatory (China-VO). This work is supported by Astronomical Big Data Joint Research Center, co-founded by National Astronomical Observatories, Chinese Academy of Sciences and Alibaba Cloud. The code and data used in this paper can be found at the PaperData Repository powered by the China-VO team. \\


\appendix

\section{Catalog of All Newly Discovered Strong Lensing System Candidates}\label{appedix}
\begin{figure*}
\centering
\includegraphics[width=0.45\textwidth]{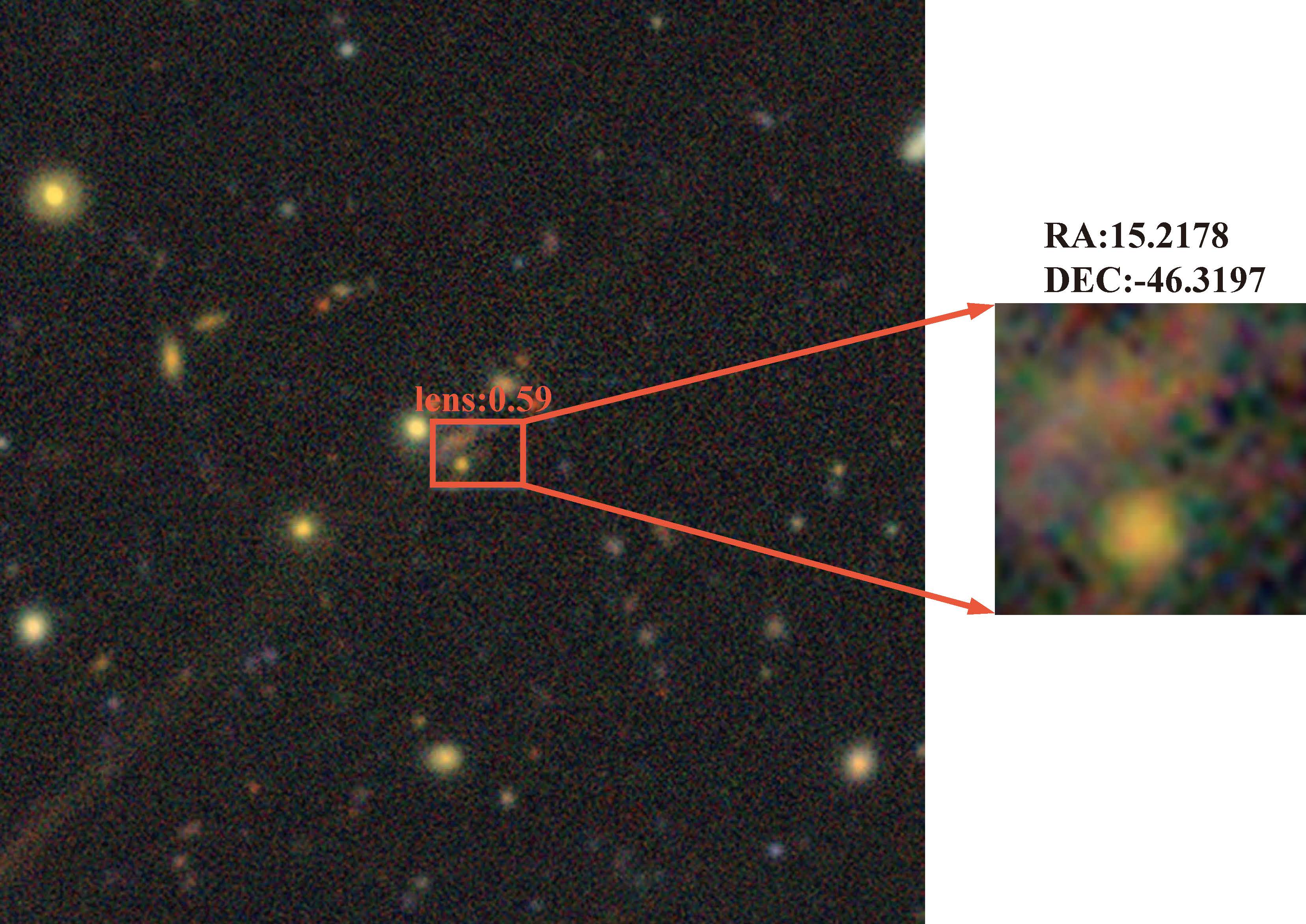}
\includegraphics[width=0.45\textwidth]{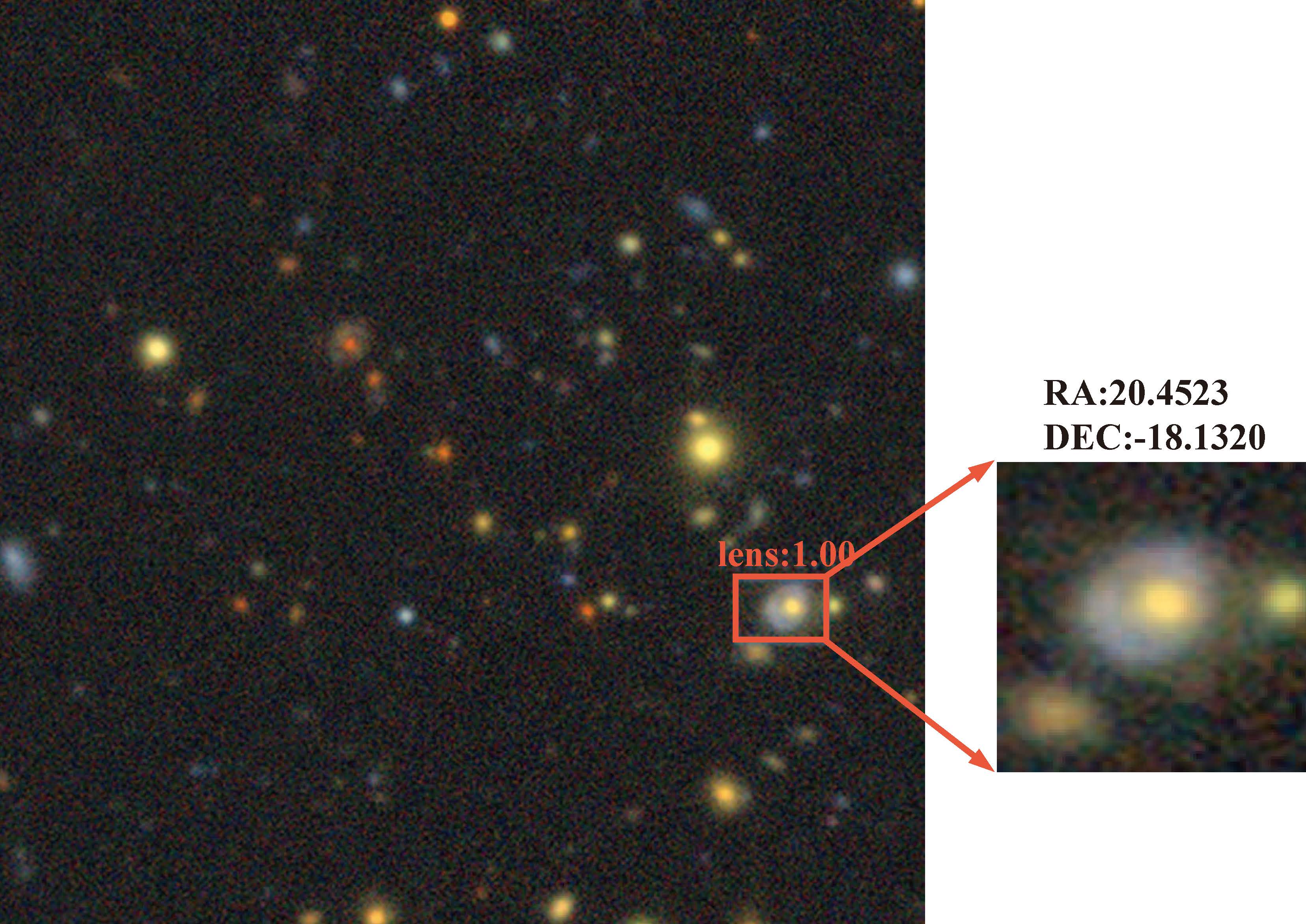}
\includegraphics[width=0.45\textwidth]{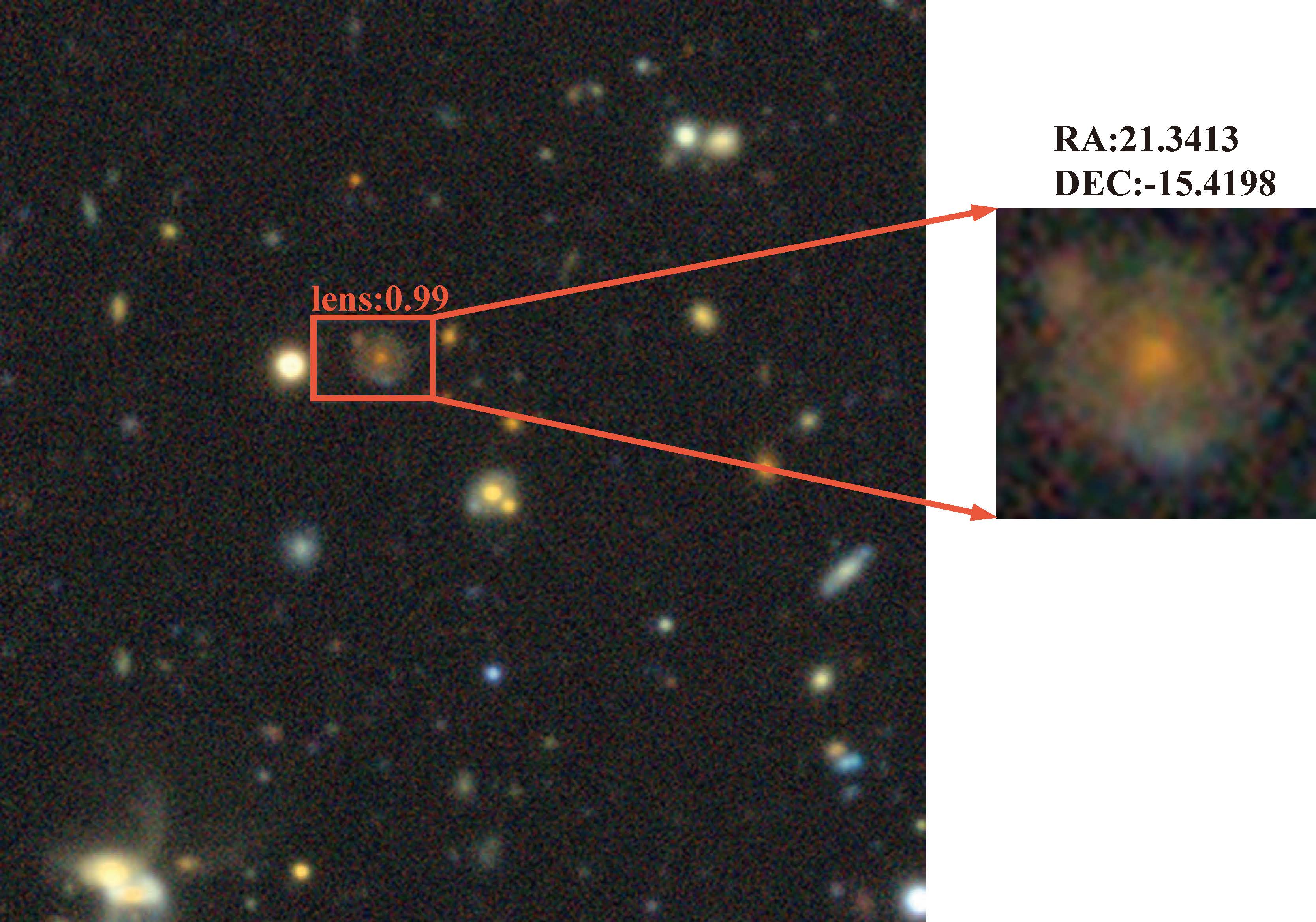}
\includegraphics[width=0.45\textwidth]{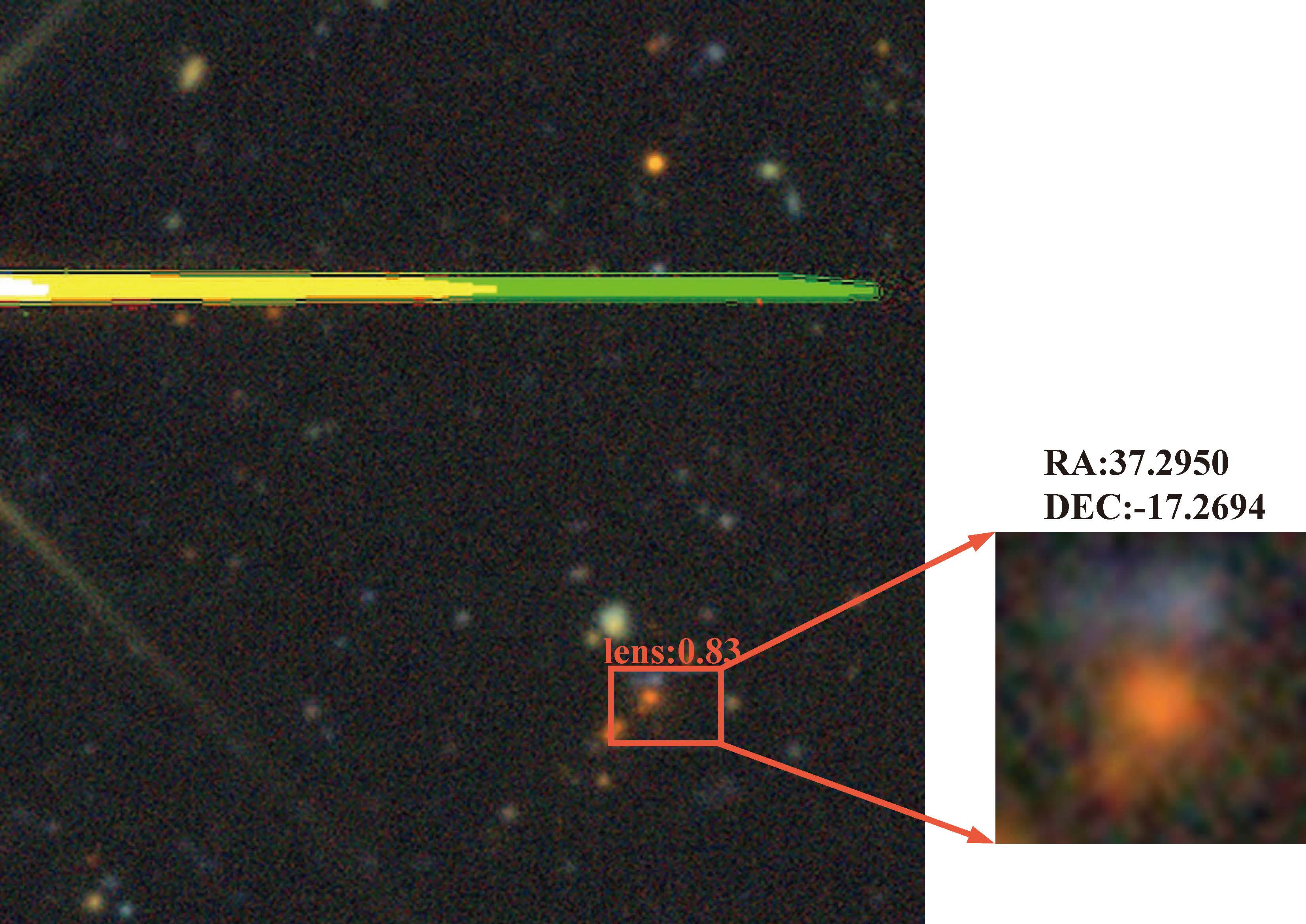}
\includegraphics[width=0.45\textwidth]{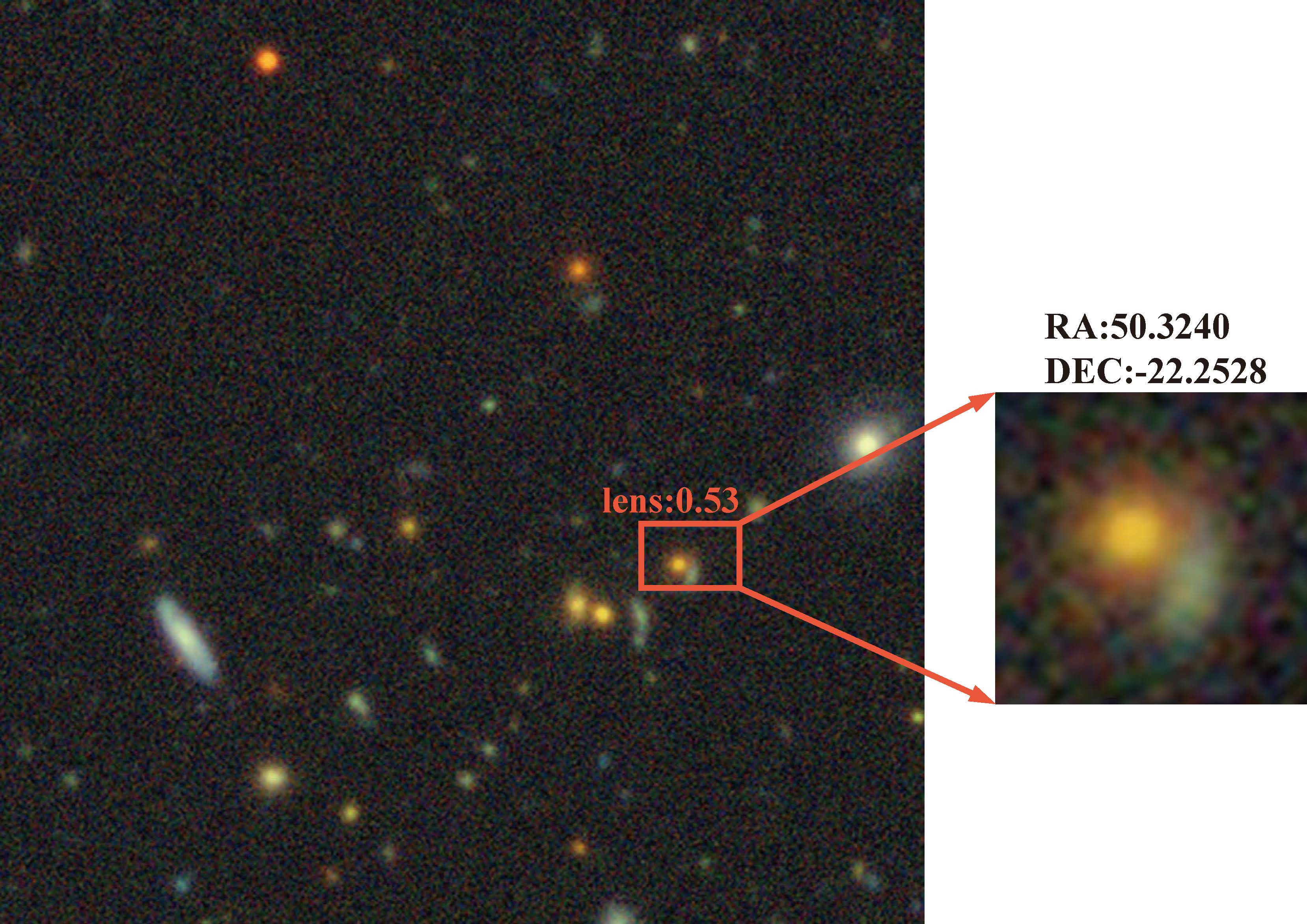}
\includegraphics[width=0.45\textwidth]{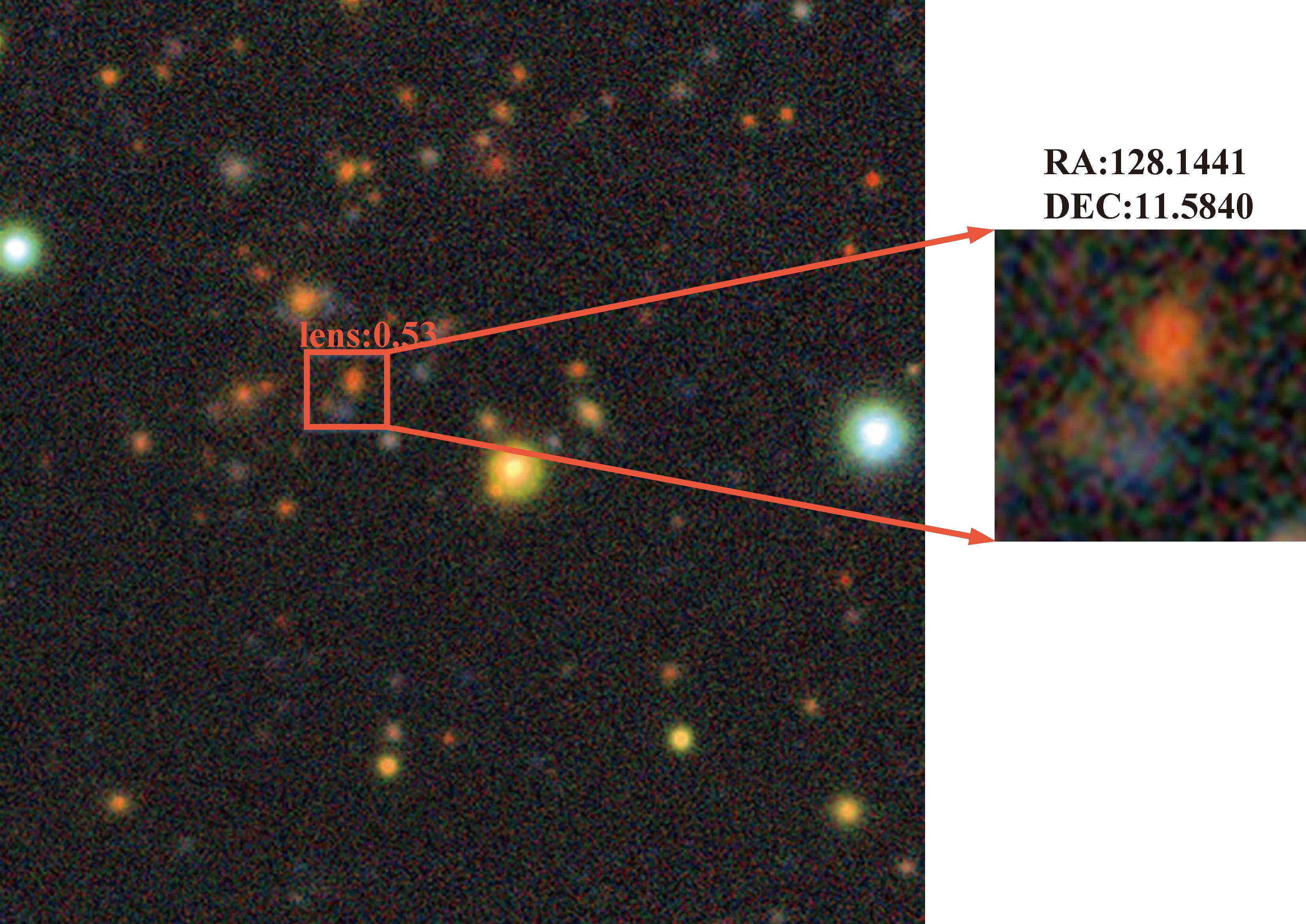}
\caption{The figure demonstrates several detection candidates obtained from real observation data by our method. It can be observed that our model accurately detects and identifies strong lensing systems at arbitrary positions in the image.}
\label{Fig15}
\end{figure*}

In this section, we have presented an overview of all strong lensing system candidates in figure \ref{Fig15}, complete with their positions and images. As illustrated in the figure, it is noteworthy that while the majority of strong lensing systems have been initially identified by scientists, our method has managed to discover additional candidates. Many of these candidates exhibit challenges such as low signal-to-noise ratios or interference from the intense illumination from nearby bright central galaxies, factors that might have rendered them elusive in earlier detection attempts.\\

\begin{table*}
\centering
\caption{Strong lensing system candidates obtained from some images obtained by the DESI Legacy Survey DR9. In this table, we show the RA, Dec and confidence of detection results returned by our algorithm.}
\begin{tabular}{cccc} 
\toprule
Name & RA & Dec & Confidence      \\ 
\hline 
DESI-3.7326+1.6078 & 3.7326 & 1.6078 & 0.65 \\
DESI-15.2178-46.3197 & 15.2178 & -46.3197 & 0.59 \\
DESI-19.6827-61.9314 & 19.6827 & -61.9314 & 0.08 \\
DESI-20.4523-18.132 & 20.4523 & -18.132 & 1.00 \\
DESI-21.3413-15.4198 & 21.3413 & -15.4198 & 0.99 \\
DESI-37.2950-17.2694 & 37.2950 & -17.2694 & 0.83 \\
DESI-38.1692-49.2411 & 38.1692 & -49.2411 & 0.89 \\
DESI-45.0621-39.7797 & 45.0621 & -39.7797 & 0.05 \\
DESI-50.324-22.2528 & 50.324 & -22.2528 & 0.53 \\
DESI-57.0827-36.0632 & 57.0827 & -36.0632 & 0.99 \\
DESI-65.0649-65.2484 & 65.0649 & -65.2484 & 0.58 \\
DESI-65.6016-61.3769 & 65.6016 & -61.3769 & 0.97 \\
DESI-65.7486-4.1514 & 65.7486 & -4.1514 & 0.57 \\
DESI-65.7605-4.1592 & 65.7605 & -4.1592 & 0.58 \\
DESI-73.516-51.3344 & 73.516 & -51.3344 & 0.96 \\
DESI-75.4883-41.3736 & 75.4883 & -41.3736 & 1.00\\
DESI-80.3946-44.7907 & 80.3946 & -44.7907 & 0.97 \\
DESI-81.2582-26.4191 & 81.2582 & -26.4191 & 0.05 \\
DESI-84.1911-45.3239 & 84.1911 & -45.3239 & 0.81 \\
DESI-87.4004-23.927 & 87.4004 & -23.927 & 0.78 \\
DESI-92.3782-59.4316 & 92.3782 & -59.4316 & 0.09 \\
DESI-128.1441+11.584 & 128.1441 & 11.584 & 0.53 \\
DESI-220.3862-0.9106 & 220.3862 & -0.9106 & 0.82 \\
DESI-355.2487-39.8141 & 355.2487 & -39.8141 & 0.11 \\
\bottomrule
\end{tabular}
\label{table3}
\end{table*}

\newpage

\section{Catalog of Detected Strong Lensing System Candidates from Euclid Early Released observation Images} \label{appedixb}
\begin{figure*}
    \foreach \i in {1,...,25} {
        \begin{subfigure}{0.19\textwidth}
            \centering
            \includegraphics[width=\linewidth]{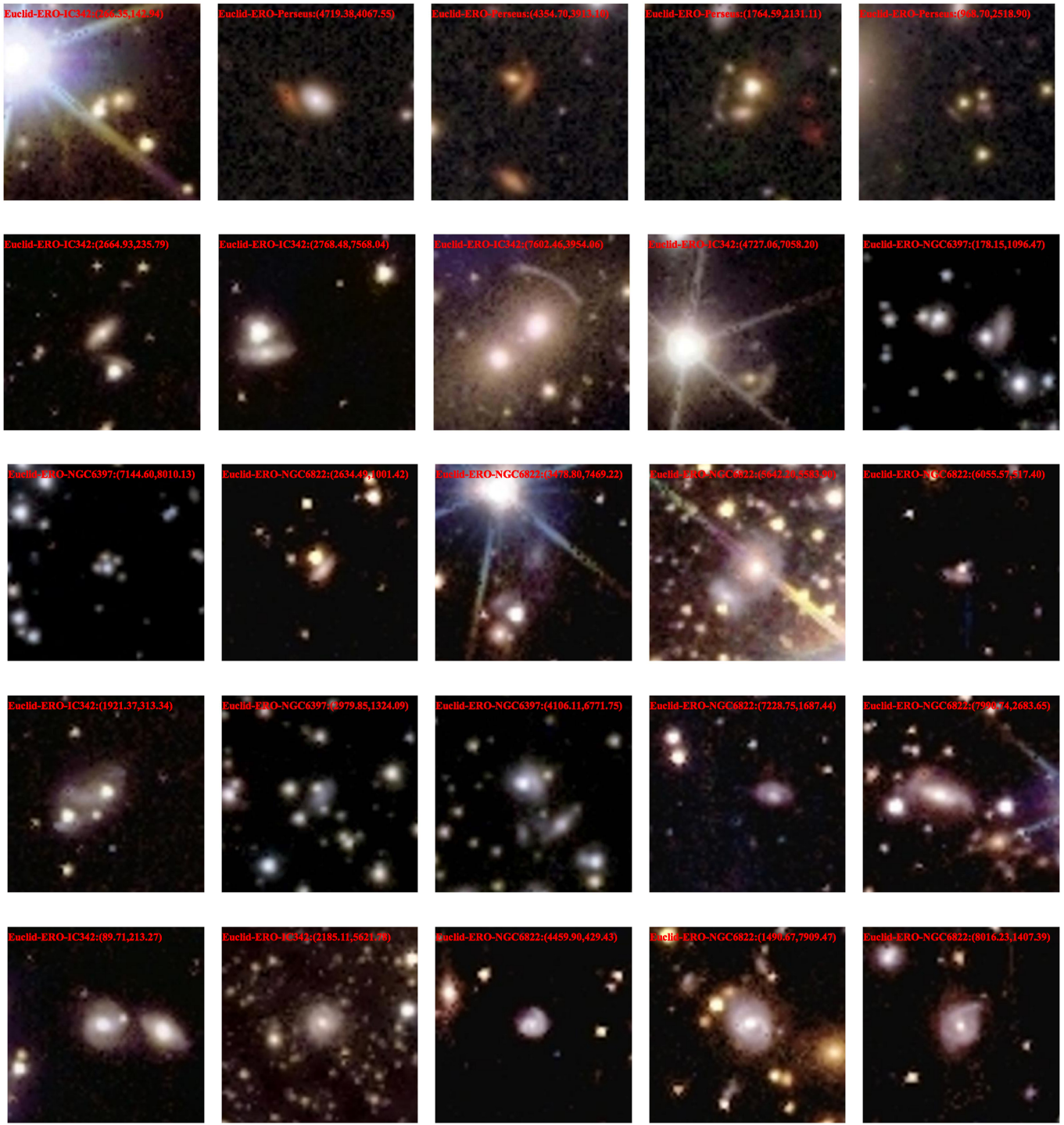}
            \label{fig:figure16.\i}
        \end{subfigure}    
        \ifnum\i=5\par\else\vspace{0.5pt}\fi 
    }
    \caption{The above figure shows the results of our method in identifying strong lensing systems in Euclid media images. We have individually examined and categorized all these images based on their morphology, classifying them into strong lensing systems of varying quality. The top three rows exhibit high-quality instances of strong lensing systems, whereas the fourth row displays strong lensing systems of moderate quality. The fifth row reveals some false positive detection results. The coordinates of these identified targets in different Euclid media images are highlighted in red color.}
    \label{Fig16}
\end{figure*}

On 7 November 2023, the European Space Agency released five public images \footnote{\url{https://www.esa.int/Science_Exploration/Space_Science/Euclid/Euclid_s_first_images_the_dazzling_edge_of_darkness}}. We have chosen four of them to evaluate the performance of our algorithm, excluding images of the Horse Head Nebula because of their complex background. Using the deep neural network with weights trained on CSST simulation data, we directly detect strong lensing systems in these images. Subsequently, we visually inspect and classify the results into strong lensing systems with high quality, those with moderate qualities, and false positives. All strong lensing systems with high quality and some with moderate quality are shown in Figure~\ref{Fig16}. The figure illustrates that our method effectively identifies strong lensing systems. Additionally, we show several typical false detection results in Figure~\ref{Fig16}. These false positives are primarily caused by diffraction spikes and spiral galaxies, which share structural similarities with arcs in strong lensing systems and are not included in the training data.\\

\newpage %

\bibliography{aj}{}

\begin{thebibliography}{}
\expandafter\ifx\csname natexlab\endcsname\relax\def\natexlab#1{#1}\fi
\providecommand{\url}[1]{\href{#1}{#1}}
\providecommand{\dodoi}[1]{doi:~\href{http://doi.org/#1}{\nolinkurl{#1}}}
\providecommand{\doeprint}[1]{\href{http://ascl.net/#1}{\nolinkurl{http://ascl.net/#1}}}
\providecommand{\doarXiv}[1]{\href{https://arxiv.org/abs/#1}{\nolinkurl{https://arxiv.org/abs/#1}}}

\bibitem[{Alard(2006)}]{alard2006automated}
Alard, C. 2006, arXiv preprint astro-ph/0606757

\bibitem[{Auger {et~al.}(2010)Auger, Treu, Bolton, Gavazzi, Koopmans, Marshall,
  Moustakas, \& Burles}]{auger2010sloan}
Auger, M., Treu, T., Bolton, A., {et~al.} 2010, The Astrophysical Journal, 724,
  511

\bibitem[{Bacon {et~al.}(2000)Bacon, Refregier, \& Ellis}]{bacon2000detection}
Bacon, D.~J., Refregier, A.~R., \& Ellis, R.~S. 2000, Monthly Notices of the
  Royal Astronomical Society, 318, 625

\bibitem[{Birrer {et~al.}(2022)Birrer, Millon, Sluse, Shajib, Courbin,
  Koopmans, Suyu, \& Treu}]{birrer2022time}
Birrer, S., Millon, M., Sluse, D., {et~al.} 2022, arXiv preprint
  arXiv:2210.10833

\bibitem[{Brada{\v{c}} {et~al.}(2002)Brada{\v{c}}, Schneider, Steinmetz,
  Lombardi, King, \& Porcas}]{bradavc2002b1422+}
Brada{\v{c}}, M., Schneider, P., Steinmetz, M., {et~al.} 2002, Astronomy \&
  Astrophysics, 388, 373

\bibitem[{Brault \& Gavazzi(2015)}]{brault2015extensive}
Brault, F., \& Gavazzi, R. 2015, Astronomy \& Astrophysics, 577, A85

\bibitem[{Chan {et~al.}(2015)Chan, Suyu, Chiueh, More, Marshall, Coupon, Oguri,
  \& Price}]{chan2015chitah}
Chan, J.~H., Suyu, S.~H., Chiueh, T., {et~al.} 2015, The Astrophysical Journal,
  807, 138

\bibitem[{Courbin {et~al.}(2000)Courbin, Lidman, Meylan, Kneib, \&
  Magain}]{courbin2000exploring}
Courbin, F., Lidman, C., Meylan, G., Kneib, J.-P., \& Magain, P. 2000, arXiv
  preprint astro-ph/0006168

\bibitem[{Dieleman {et~al.}(2015)Dieleman, Willett, \&
  Dambre}]{dieleman2015rotation}
Dieleman, S., Willett, K.~W., \& Dambre, J. 2015, Monthly notices of the royal
  astronomical society, 450, 1441

\bibitem[{Dosovitskiy {et~al.}(2020)Dosovitskiy, Beyer, Kolesnikov,
  Weissenborn, Zhai, Unterthiner, Dehghani, Minderer, Heigold, Gelly,
  {et~al.}}]{dosovitskiy2020image}
Dosovitskiy, A., Beyer, L., Kolesnikov, A., {et~al.} 2020, arXiv preprint
  arXiv:2010.11929

\bibitem[{Estrada {et~al.}(2007)Estrada, Annis, Diehl, Hall, Las, Lin, Makler,
  Merritt, Scarpine, Allam, {et~al.}}]{estrada2007systematic}
Estrada, J., Annis, J., Diehl, H., {et~al.} 2007, The Astrophysical Journal,
  660, 1176

\bibitem[{{Faber} {et~al.}(2007){Faber}, {Willmer}, {Wolf}, {Koo}, {Weiner},
  {Newman}, {Im}, {Coil}, {Conroy}, {Cooper}, {Davis}, {Finkbeiner}, {Gerke},
  {Gebhardt}, {Groth}, {Guhathakurta}, {Harker}, {Kaiser}, {Kassin},
  {Kleinheinrich}, {Konidaris}, {Kron}, {Lin}, {Luppino}, {Madgwick},
  {Meisenheimer}, {Noeske}, {Phillips}, {Sarajedini}, {Schiavon}, {Simard},
  {Szalay}, {Vogt}, \& {Yan}}]{Faber2007}
{Faber}, S.~M., {Willmer}, C.~N.~A., {Wolf}, C., {et~al.} 2007, \apj, 665, 265,
  \dodoi{10.1086/519294}

\bibitem[{Fortson {et~al.}(2012)Fortson, Masters, Nichol, Edmondson, Lintott,
  Raddick, \& Wallin}]{fortson2012galaxy}
Fortson, L., Masters, K., Nichol, R., {et~al.} 2012, Advances in machine
  learning and data mining for astronomy, 2012, 213

\bibitem[{Gavazzi {et~al.}(2014)Gavazzi, Marshall, Treu, \&
  Sonnenfeld}]{gavazzi2014ringfinder}
Gavazzi, R., Marshall, P.~J., Treu, T., \& Sonnenfeld, A. 2014, The
  Astrophysical Journal, 785, 144

\bibitem[{Golse \& Kneib(2002)}]{golse2002pseudo}
Golse, G., \& Kneib, J.-P. 2002, Astronomy \& Astrophysics, 390, 821

\bibitem[{Goodfellow {et~al.}(2014)Goodfellow, Pouget-Abadie, Mirza, Xu,
  Warde-Farley, Ozair, Courville, \& Bengio}]{goodfellow2014generative}
Goodfellow, I., Pouget-Abadie, J., Mirza, M., {et~al.} 2014, Advances in neural
  information processing systems, 27

\bibitem[{Huang {et~al.}(2020)Huang, Storfer, Ravi, Pilon, Domingo, Schlegel,
  Bailey, Dey, Gupta, Herrera, {et~al.}}]{huang2020finding}
Huang, X., Storfer, C., Ravi, V., {et~al.} 2020, The Astrophysical Journal,
  894, 78

\bibitem[{Ivezi{\'c} {et~al.}(2019)Ivezi{\'c}, Kahn, Tyson, Abel, Acosta,
  Allsman, Alonso, AlSayyad, Anderson, Andrew, {et~al.}}]{ivezic2019lsst}
Ivezi{\'c}, {\v{Z}}., Kahn, S.~M., Tyson, J.~A., {et~al.} 2019, The
  Astrophysical Journal, 873, 111

\bibitem[{Jacobs {et~al.}(2017)Jacobs, Glazebrook, Collett, More, \&
  McCarthy}]{jacobs2017finding}
Jacobs, C., Glazebrook, K., Collett, T., More, A., \& McCarthy, C. 2017,
  Monthly Notices of the Royal Astronomical Society, 471, 167

\bibitem[{Jia {et~al.}(2023{\natexlab{a}})Jia, Liu, Liu, \&
  Pan}]{jia2023target}
Jia, P., Liu, W., Liu, Y., \& Pan, H. 2023{\natexlab{a}}, The Astrophysical
  Journal Supplement Series, 264, 43

\bibitem[{Jia {et~al.}(2024)Jia, Lv, Ning, Song, Li, Ji, Cui, \&
  Li}]{jia2024image}
Jia, P., Lv, J., Ning, R., {et~al.} 2024, Monthly Notices of the Royal
  Astronomical Society, 527, 6581

\bibitem[{Jia {et~al.}(2022)Jia, Sun, Li, Song, Ning, Wei, \&
  Luo}]{jia2022detection}
Jia, P., Sun, R., Li, N., {et~al.} 2022, The Astronomical Journal, 165, 26

\bibitem[{Jia {et~al.}(2020)Jia, Wu, Yi, Cai, \& Cai}]{jia2020psf}
Jia, P., Wu, X., Yi, H., Cai, B., \& Cai, D. 2020, The Astronomical Journal,
  159, 183

\bibitem[{Jia {et~al.}(2023{\natexlab{b}})Jia, Zheng, Wang, \&
  Yang}]{jia2023deep}
Jia, P., Zheng, Y., Wang, M., \& Yang, Z. 2023{\natexlab{b}}, Astronomy and
  Computing, 42, 100687

\bibitem[{Jiang {et~al.}(2021)Jiang, Dai, Wu, \& Loy}]{jiang2021focal}
Jiang, L., Dai, B., Wu, W., \& Loy, C.~C. 2021, in Proceedings of the IEEE/CVF
  International Conference on Computer Vision, 13919--13929

\bibitem[{{Keeton}(2001)}]{Keeton2001}
{Keeton}, C.~R. 2001, arXiv e-prints, astro,
  \dodoi{10.48550/arXiv.astro-ph/0102341}

\bibitem[{Kneib \& Natarajan(2011)}]{kneib2011cluster}
Kneib, J.-P., \& Natarajan, P. 2011, The Astronomy and Astrophysics Review, 19,
  1

\bibitem[{{Kormann} {et~al.}(1994){Kormann}, {Schneider}, \&
  {Bartelmann}}]{Kormann1994}
{Kormann}, R., {Schneider}, P., \& {Bartelmann}, M. 1994, \aap, 284, 285

\bibitem[{{Korytov} {et~al.}(2019){Korytov}, {Hearin}, {Kovacs}, {Larsen},
  {Rangel}, {Hollowed}, {Benson}, {Heitmann}, {Mao}, {Bahmanyar}, {Chang},
  {Campbell}, {DeRose}, {Finkel}, {Frontiere}, {Gawiser}, {Habib}, {Joachimi},
  {Lanusse}, {Li}, {Mandelbaum}, {Morrison}, {Newman}, {Pope}, {Rykoff},
  {Simet}, {To}, {Vikraman}, {Wechsler}, {White}, \& {(The LSST Dark Energy
  Science Collaboration}}]{2019Korytov}
{Korytov}, D., {Hearin}, A., {Kovacs}, E., {et~al.} 2019, \apjs, 245, 26,
  \dodoi{10.3847/1538-4365/ab510c}

\bibitem[{Laureijs {et~al.}(2011)Laureijs, Amiaux, Arduini, Augueres,
  Brinchmann, Cole, Cropper, Dabin, Duvet, Ealet, \&
  others.}]{2011arXiv1110.3193L}
Laureijs, R., Amiaux, J., Arduini, S., {et~al.} 2011, arXiv e-prints,
  arXiv:1110.3193, \dodoi{10.48550/arXiv.1110.3193}

\bibitem[{{Lauritsen} {et~al.}(2021){Lauritsen}, {Dickinson}, {Bromley},
  {Serjeant}, {Lim}, {Gao}, \& {Wang}}]{Lauritsen2021}
{Lauritsen}, L., {Dickinson}, H., {Bromley}, J., {et~al.} 2021, \mnras, 507,
  1546, \dodoi{10.1093/mnras/stab2195}

\bibitem[{Lenzen {et~al.}(2004)Lenzen, Schindler, \&
  Scherzer}]{lenzen2004automatic}
Lenzen, F., Schindler, S., \& Scherzer, O. 2004, Astronomy \& Astrophysics,
  416, 391

\bibitem[{Li {et~al.}(2016)Li, Gladders, Rangel, Florian, Bleem, Heitmann,
  Habib, \& Fasel}]{li2016pics}
Li, N., Gladders, M.~D., Rangel, E.~M., {et~al.} 2016, The Astrophysical
  Journal, 828, 54

\bibitem[{Lin {et~al.}(2017)Lin, Doll{\'a}r, Girshick, He, Hariharan, \&
  Belongie}]{lin2017feature}
Lin, T.-Y., Doll{\'a}r, P., Girshick, R., {et~al.} 2017, in Proceedings of the
  IEEE conference on computer vision and pattern recognition, 2117--2125

\bibitem[{Liu {et~al.}(2021)Liu, Lin, Cao, Hu, Wei, Zhang, Lin, \&
  Guo}]{liu2021swin}
Liu, Z., Lin, Y., Cao, Y., {et~al.} 2021, in Proceedings of the IEEE/CVF
  international conference on computer vision, 10012--10022

\bibitem[{Lupton {et~al.}(2004)Lupton, Blanton, Fekete, Hogg, O’Mullane,
  Szalay, \& Wherry}]{lupton2004preparing}
Lupton, R., Blanton, M.~R., Fekete, G., {et~al.} 2004, Publications of the
  Astronomical Society of the Pacific, 116, 133

\bibitem[{Lv {et~al.}(2022)Lv, Ning, Song, \& Jia}]{lv2022general}
Lv, J., Ning, R., Song, Y., \& Jia, P. 2022, in Software and
  Cyberinfrastructure for Astronomy VII, Vol. 12189, SPIE, 694--703

\bibitem[{{Madireddy} {et~al.}(2019){Madireddy}, {Li}, {Ramachandra}, {Butler},
  {Balaprakash}, {Habib}, \& {Heitmann}}]{Madireddy2019}
{Madireddy}, S., {Li}, N., {Ramachandra}, N., {et~al.} 2019, arXiv e-prints,
  arXiv:1911.03867.
\newblock \doarXiv{1911.03867}

\bibitem[{Meneghetti {et~al.}(2013)Meneghetti, Bartelmann, Dahle, \&
  Limousin}]{meneghetti2013arc}
Meneghetti, M., Bartelmann, M., Dahle, H., \& Limousin, M. 2013, Space Science
  Reviews, 177, 31

\bibitem[{Metcalf {et~al.}(2019)Metcalf, Meneghetti, Avestruz, Bellagamba, Bom,
  Bertin, Cabanac, Courbin, Davies, Decenci{\`e}re,
  {et~al.}}]{metcalf2019strong}
Metcalf, R.~B., Meneghetti, M., Avestruz, C., {et~al.} 2019, Astronomy \&
  Astrophysics, 625, A119

\bibitem[{More {et~al.}(2012)More, Cabanac, More, Alard, Limousin, Kneib,
  Gavazzi, \& Motta}]{more2012cfhtls}
More, A., Cabanac, R., More, S., {et~al.} 2012, The Astrophysical Journal, 749,
  38

\bibitem[{More {et~al.}(2016)More, Verma, Marshall, More, Baeten, Wilcox,
  Macmillan, Cornen, Kapadia, Parrish, {et~al.}}]{more2016space}
More, A., Verma, A., Marshall, P.~J., {et~al.} 2016, Monthly Notices of the
  Royal Astronomical Society, 455, 1191

\bibitem[{{Navarro} {et~al.}(1996){Navarro}, {Frenk}, \& {White}}]{NFW1996}
{Navarro}, J.~F., {Frenk}, C.~S., \& {White}, S. D.~M. 1996, \apj, 462, 563,
  \dodoi{10.1086/177173}

\bibitem[{Oguri \& Marshall(2010)}]{oguri2010gravitationally}
Oguri, M., \& Marshall, P.~J. 2010, Monthly Notices of the Royal Astronomical
  Society, 405, 2579

\bibitem[{Parker {et~al.}(2007)Parker, Hoekstra, Hudson, Van~Waerbeke, \&
  Mellier}]{parker2007masses}
Parker, L.~C., Hoekstra, H., Hudson, M.~J., Van~Waerbeke, L., \& Mellier, Y.
  2007, The Astrophysical Journal, 669, 21

\bibitem[{Petrillo {et~al.}(2017)Petrillo, Tortora, Chatterjee, Vernardos,
  Koopmans, Verdoes~Kleijn, Napolitano, Covone, Schneider, Grado,
  {et~al.}}]{petrillo2017finding}
Petrillo, C.~E., Tortora, C., Chatterjee, S., {et~al.} 2017, Monthly Notices of
  the Royal Astronomical Society, 472, 1129

\bibitem[{Ronneberger {et~al.}(2015)Ronneberger, Fischer, \&
  Brox}]{ronneberger2015u}
Ronneberger, O., Fischer, P., \& Brox, T. 2015, in Medical Image Computing and
  Computer-Assisted Intervention--MICCAI 2015: 18th International Conference,
  Munich, Germany, October 5-9, 2015, Proceedings, Part III 18, Springer,
  234--241

\bibitem[{Rowe {et~al.}(2015)Rowe, Jarvis, Mandelbaum, Bernstein, Bosch, Simet,
  Meyers, Kacprzak, Nakajima, Zuntz, {et~al.}}]{rowe2015galsim}
Rowe, B.~T., Jarvis, M., Mandelbaum, R., {et~al.} 2015, Astronomy and
  Computing, 10, 121

\bibitem[{Schawinski {et~al.}(2017)Schawinski, Zhang, Zhang, Fowler, \&
  Santhanam}]{schawinski2017generative}
Schawinski, K., Zhang, C., Zhang, H., Fowler, L., \& Santhanam, G.~K. 2017,
  Monthly Notices of the Royal Astronomical Society: Letters, 467, L110

\bibitem[{Seidel \& Bartelmann(2007)}]{seidel2007arcfinder}
Seidel, G., \& Bartelmann, M. 2007, Astronomy \& Astrophysics, 472, 341

\bibitem[{Shajib {et~al.}(2022)Shajib, Vernardos, Collett, Motta, Sluse,
  Williams, Saha, Birrer, Spiniello, \& Treu}]{shajib2022strong}
Shajib, A., Vernardos, G., Collett, T., {et~al.} 2022, arXiv preprint
  arXiv:2210.10790

\bibitem[{Smith {et~al.}(2001)Smith, Kneib, Ebeling, Czoske, \&
  Smail}]{smith2001hubble}
Smith, G.~P., Kneib, J.-P., Ebeling, H., Czoske, O., \& Smail, I. 2001, The
  Astrophysical Journal, 552, 493

\bibitem[{Sonnenfeld {et~al.}(2015)Sonnenfeld, Treu, Marshall, Suyu, Gavazzi,
  Auger, \& Nipoti}]{sonnenfeld2015sl2s}
Sonnenfeld, A., Treu, T., Marshall, P.~J., {et~al.} 2015, The Astrophysical
  Journal, 800, 94

\bibitem[{Treu(2010)}]{treu2010strong}
Treu, T. 2010, Annual Review of Astronomy and Astrophysics, 48, 87

\bibitem[{Treu \& Koopmans(2002)}]{treu2002internal}
Treu, T., \& Koopmans, L.~V. 2002, The Astrophysical Journal, 575, 87

\bibitem[{Treu {et~al.}(2022)Treu, Suyu, \& Marshall}]{treu2022strong}
Treu, T., Suyu, S.~H., \& Marshall, P.~J. 2022, The Astronomy and Astrophysics
  Review, 30, 8

\bibitem[{Vaswani {et~al.}(2017)Vaswani, Shazeer, Parmar, Uszkoreit, Jones,
  Gomez, Kaiser, \& Polosukhin}]{vaswani2017attention}
Vaswani, A., Shazeer, N., Parmar, N., {et~al.} 2017, Advances in neural
  information processing systems, 30

\bibitem[{Vegetti {et~al.}(2023)Vegetti, Birrer, Despali, Fassnacht, Gilman,
  Hezaveh, Levasseur, McKean, Powell, O'Riordan, {et~al.}}]{vegetti2023strong}
Vegetti, S., Birrer, S., Despali, G., {et~al.} 2023, arXiv preprint
  arXiv:2306.11781

\bibitem[{Wang {et~al.}(2004)Wang, Bovik, Sheikh, \&
  Simoncelli}]{wang2004image}
Wang, Z., Bovik, A.~C., Sheikh, H.~R., \& Simoncelli, E.~P. 2004, IEEE
  transactions on image processing, 13, 600

\bibitem[{Webster {et~al.}(1988)Webster, Hewett, \&
  Irwin}]{webster1988automated}
Webster, R.~L., Hewett, P.~C., \& Irwin, M.~J. 1988, Astronomical Journal (ISSN
  0004-6256), vol. 95, Jan. 1988, p. 19-25. Research supported by Cambridge
  University., 95, 19

\bibitem[{Xu {et~al.}(2014)Xu, Ren, Liu, \& Jia}]{xu2014deep}
Xu, L., Ren, J.~S., Liu, C., \& Jia, J. 2014, Advances in neural information
  processing systems, 27

\bibitem[{Zhan(2021)}]{CSSTHu}
Zhan, H. 2021, Chinese Science Bulletin, 66, 1290

\end{thebibliography}
\bibliographystyle{aasjournal}

\end{document}